\documentclass[12pt]{iopart}

\usepackage{dcolumn}
\usepackage{xcolor}
\usepackage[table]{colortbl}
\usepackage{multirow}
\usepackage{bm}
\usepackage{array,hhline} 

\usepackage{amssymb}
\usepackage{graphicx}
\usepackage{multicol}

\newcommand{\vect}[1]{\mbox{\boldmath $#1$}}

\definecolor{Silver}{rgb}{0.9,0.9,0.9}

\begin{document}

\title{Orthogonal versus covariant Lyapunov vectors for rough hard disk systems}

\author{H.~Bosetti\footnote{Present address: Section for Science of Complex Systems, Medical University of Vienna, Spitalgasse 23, 1090 Wien, Austria}
 and H.A.~Posch}

\address{Computational Physics Group, Faculty of Physics, University of Vienna, Boltzmanngasse 5, 1090 Wien, Austria}
\ead{hadrien.bosetti@univie.ac.at}
\ead{harald.posch@univie.ac.at}

\begin{abstract}
The Oseledec splitting of the tangent space into covariant subspaces for a hyperbolic dynamical system
is  numerically accessible by computing the full set of covariant Lyapunov vectors.
In this paper, the covariant Lyapunov vectors, the orthogonal Gram-Schmidt vectors, and the corresponding local (time-dependent)
Lyapunov exponents, are analyzed for a planar system of rough hard disks (RHDS).
These results are compared to respective results for a smooth-hard-disk system (SHDS).
We find that the rotation of the disks deeply affects the Oseledec splitting and the structure of the tangent space.
For both the smooth and rough hard disks, the stable, unstable and central manifolds
are transverse to each other, although the minimal angle between the unstable and stable manifolds of the RHDS typically is very small.
Both systems are hyperbolic. However, the central manifold is precisely orthogonal to the rest 
of the tangent space only for the  smooth-particle case and not for the rough disks. We also demonstrate  
that the rotations destroy the Hamiltonian  character for the  rough-hard-disk system.
\end{abstract}
\maketitle

\section{Introduction}\label{rough_introduction}

The phase-space trajectory of  particle systems generally is sensitive to tiny (infinitesimal) perturbations.
The ensuing exponential instability is  revealed by following the time evolution of a complete set of  perturbation vectors,
$\lbrace \delta {\bf \Gamma}_i \rbrace_{i=1}^{\mathtt{D}}$,  spanning the tangent space, and
by computing the whole set of exponential rate constants, the Lyapunov exponents. Here,  $\mathtt{D}$ denotes the phase-space dimension.
Two kinds of perturbation vectors are usually employed: the orthonormal Gram-Schmidt vectors, $\lbrace\vect{g}_i \rbrace_{i=1}^{\mathtt{D}}$,
and the generally non-orthogonal covariant Lyapunov vectors, $\{ \vect{v}_i \}_{i=1}^{\mathtt{D}}$. The former were introduced  
in 1980 by Benettin \textit{et al.} \cite{Benettin1:1980,Benettin2:1980} and by Shimada and Nagashima \cite{Shimada:1979} for the computation of the 
spectrum of Lyapunov exponents. They  are made  orthonormal  by a periodic applications of  a Gram-Schmidt  procedure
to counteract  the tendency of all perturbation vectors to rotate towards the most unstable direction in phase space.
This procedure is equivalent to a QR decomposition at periodic time intervals \cite{Eckmann:1985, Johnson:1987}.
For a recent review we refer to Ref. \cite{Skokos}.
The covariant vectors, on the other hand, are generally non-orthogonal and were made numerically accessible in 2007 by Ginelli \textit{et al.} \cite{Ginelli}
and by Wolfe and Samelson \cite{Wolfe:2007}. They co-move (co-reorient, in particular)  with the flow in tangent space and define a hierarchy of directions 
at any phase-space point, which are  individually associated with the Lyapunov exponents (as long as the exponents are non-degenerate). 
The algorithm for their construction is more involved and still requires the computation of the Gram-Schmidt vectors in the forward direction of time,
followed by an iteration in the backward direction to obtain the covariant vectors (see \cite{Hadrien2010_1,Kuptsov:2011}).
Both vector sets are intimately related to various subspaces of the tangent space introduced by Oseledec in 1968 \cite{Oseledec:1968}.
The Gram-Schmidt vectors are related to the eigenvectors of the Oseledec matrix \cite{ Johnson:1987,Ershov}.
The covariant vectors correspond to the Oseledec splitting \cite{Ruelle:1979,Eckmann:2005,Hadrien2010_1}
and, thus, provide a very natural spanning set of the tangent space at any phase point.

The time-averaged growth rate of a covariant vector gives rise to a (global) Lyapunov exponent,
and the set of exponents, $\lbrace \lambda_i \rbrace_{i=1}^{\mathtt{D}}$, ordered according to size, $\lambda_i > \lambda_{i+1}$,  is referred to as the Lyapunov spectrum \cite{Oseledec:1968}.
The sub-exponential growth rates give vanishing exponents, for which the associated covariant vectors span the central manifold.
The covariant vectors with positive (negative) Lyapunov exponents span the unstable (stable) manifold.
For conservative systems the sum of all Lyapunov exponents vanishes. This sum  is negative for dissipative systems, which implies a collapse of the phase-space distribution onto a multifractal strange attractor \cite{PhysRevLett.75.2831}.
At any phase-space point  local Lyapunov exponents may be defined. They vary with time when the system evolves.
The global  exponents are time averages of the local Lyapunov exponents
\cite{Hadrien2010_2,Ramaswamy}.
 
   In this paper we study the Lyapunov instability for rough-hard-disk systems (RHDS) \cite{vMP2009}.
By adding the notion of roughness  and angular momentum to each particle,  
an exchange of energy between translational and rotational degrees of freedom is introduced. 
The three-dimensional version of this model -- arguably -- may be considered as the simplest model of
a molecular fluid. Our results are compared to those for smooth-hard-disk systems (SHDS),
which have been the subject of many previous studies  \cite{DPH1996,Posch:2000,FHPH2004,TM2003a,TM2003b}.

   The paper is organized as follows. The basic concepts of the
dynamics of the tangent-space bundle are summarized in Sec.~\ref{tangent_bundle}.
The smooth-hard-disk system and, more thoroughly, the rough-hard-disk system
are introduced in Sec.~\ref{hard_disks}.
In Sec.~\ref{gs_covariant}  we compare the Gram-Schmidt vectors to the covariant Lyapunov vectors
(which give rise to the same global Lyapunov exponents) for the RHDS.
Our aim is to characterize the transition from non-rotating (smooth) to rotating (rough) particles.
The null subspace for the RHDS is the topic of Sec.~\ref{null_subspace},
where we show that it is not strictly orthogonal to the unstable and stable subspaces.
Using the covariant vector set, we show that the dynamics is converging to a sub-manifold.
In Sec.~\ref{transversality} we examine the loss of isolation for the local covariant exponents,
which implies that a finite-time average of local exponents may lead to a different order in the Lyapunov spectrum as compared
to that of globally-averaged exponents.
Furthermore, the transversality between the stable, the unstable, and the central manifolds is investigated for the SHDS and the RHDS.
Sec.~\ref{discussion} is devoted to a discussion of the qualitative differences between the two models and
of the break down of the Hamiltonian character for the RHDS due to the coupling between the translational and rotational 
degrees of freedom. We conclude in Sec. \ref{summary} with a short summary. 

\section{Tangent-bundle dynamics}\label{tangent_bundle}

First we recall some basic notions in statistical mechanics and dynamical systems theory
(see e.g. \cite{Arnold_2}):
\textit{(i)} The concept of a phase space $\mathbf{X}$
is tantamount to consider the full ensemble of accessible system states,
denoted by the state vector ${\bf \Gamma}$.
If this system is ergodic, which we assume, every infinitesimal volume element of the phase space
in accord with the conserved quantities will be visited in due time.
Each phase point ${\bf \Gamma}$ on the hypersurface, defined by the exclusion of the subspace violating the conservation laws, 
becomes equally likely;
\textit{(ii)} The dynamics in the phase space is characterized by the motion of a  phase point in that space
with a fixed frame,  which is not the case for the tangent space $\mathbf{TX}_{\bf \Gamma}$.
$\mathbf{TX}_{\bf \Gamma}$ moves with the state vector in phase space and is tangent to the hypersurface determined
by the conserved quantities. It contains the complete set of perturbation vectors $\lbrace\delta {\bf \Gamma}_i \rbrace$
which point in all accessible directions, where the state ${\bf \Gamma}$ is allowed to evolve
(observing the conservation rules);
\textit{(iii)} In order to examine also the dynamics of the perturbation vectors $\delta {\bf \Gamma}$,
we have to consider the concept of the tangent bundle $\mathbf{TX}$, which consists of the union of the tangent spaces to
$\mathbf{X}$
at various state vectors~${\bf \Gamma}$, $\bigcup_{{\bf \Gamma} \, \in \, \mathbf{X}}\mathbf{TX}_{\bf \Gamma}$.
A point of $\mathbf{TX}$ is a vector $\delta {\bf \Gamma}$, tangent to $\mathbf{X}$ at ${\bf \Gamma}(t)$.

According to the \textit{multiplicative ergodic theorem} of Oseledec for unstable systems \cite{Oseledec:1968},
the dynamics of the tangent bundle induces some hierarchically dissociated sub-bundles.
At some phase point ${\bf \Gamma}$, the subbundles correspond to a set of dissociated subspaces,
which is referred to as Oseledec splitting.
The exponential growth (or decay) of the subbundles gives rise to the Lyapunov spectrum,
whose hierarchical structure is reflected by the ordering of the exponents by size.
Ginelli \textit{et al.} \cite{Ginelli} introduced an algorithm which permits to generate numerically  the Oseledec splitting
{\em via} the covariant vectors $\lbrace \vect{v}_i \rbrace$.
This method is based upon a theoretical statement by Ruelle \cite{Ruelle:1979} and Ershov and Potapov \cite{Ershov},
\begin{equation}
\widehat{\vect{v}}_i = \Big(\vect{g}_1^{(+)} \oplus\cdots\oplus \vect{g}_i^{(+)} \Big) \cap \Big( \vect{g}_i^{(-)} \oplus\cdots\oplus \vect{g}_{\mathtt{D}}^{(-)}\Big)
\enskip ,
\label{oseledec_splitting}
\end{equation}
where $\widehat{\vect{v}}_i$ is a one-dimensional subspace of $\mathbf{TX}_{{\bf \Gamma}}$ spanned by the covariant vector $\vect{v}_i$, and
 $\lbrace \vect{g}_i \rbrace$ denotes the set of orthonormal Gram-Schmidt vectors. As usual, $\mathtt{D}$ is the phase space dimension.
All of these vectors are associated with the same state vector ${\bf \Gamma}$.
The superscript ``$+$'' refers to the forward direction of time
(i.e. the starting vector set $\lbrace\vect{g}^{(+)}_i\rbrace$ was initially placed in the far past),
whereas the superscript ``$-$'' refers to the backward direction of time
(i.e. the starting vector set $\lbrace\vect{g}^{(-)}_i\rbrace$ was initially placed in the far future).

Let  ${\bf \Gamma}(t)$ denote the state of the system at time $t$.
The flow in phase space is denoted by $\phi^t:  \mathbf{X} \to \mathbf{X}$
and the corresponding flow in tangent space by $D\phi^t:  \mathbf{TX}_{{\bf \Gamma}(0)} \to \mathbf{TX}_{{\bf \Gamma}(t)}$.
Then the covariant vectors $\lbrace \vect{v}_i \rbrace$ obey
\begin{displaymath}
 D \phi^t \vert_{{\bf \Gamma}(0)} \vect{v}_i({\bf \Gamma}(0)) = \vect{v}_i ({\bf \Gamma}(t))
 \enskip ,
\end{displaymath}
for all $ i \in \left\lbrace 1,\ldots, \mathtt{D} \right\rbrace$, where ${\bf \Gamma}(t) = \phi^t ({\bf \Gamma}(0))$.
The corresponding (global) Lyapunov exponents are time averages,
\begin{displaymath}
\lim_{t \rightarrow \pm \infty}
\frac{1}{\vert t \vert} \, \log \, \Vert \, D\phi^t\vert_{{\bf \Gamma}(0)}
\; \vect{v}_i ({\bf \Gamma}(0)) \, \Vert
= \pm \lambda_i
\enskip ,
\end{displaymath}
and due to the dynamical hierarchy, they obey: $\lambda_1 \geq \cdots \geq \lambda_\mathtt{D}$.
When we demonstrate  covariant vectors in our numerical work below, we shall always consider them as normalized (without 
explicit reference to the normalization procedure).
\section{Systems of rough hard disks (RHDS)}\label{hard_disks}

Rough disks temporarily store energy in internal degrees of freedom (rotation).
They suffer elastic hard collisions and move along straight lines inbetween.
The systems consist of $N$ particles at a density $\rho$ in a box of dimension $(L_x,\,L_y)$
with periodic boundaries in the $x$- and $y$-directions.
The state vector is given by
\begin{equation}
{\bf \Gamma} = \lbrace  \vec{q}_n , \,  \vec{p}_n , \omega_n \rbrace_{n=1}^N
\enspace ,
\label{state_2}
\end{equation}
where $\vec{q}_n , \,  \vec{p}_n , \omega_n$ denote the respective position, momentum and angular velocity of particle $n$.
The orientation angles of the disks are not required for the equations of motion and are not included in the list of
independent variables.
This means that the Lagrangian of the system does not depend on the orientation of the particles.
We shall come back to this point in Sec.~\ref{null_subspace}.
An arbitrary perturbation vector is given by
\begin{equation}
\delta {\bf \Gamma} = \lbrace \delta \vec{q}_n , \, \delta \vec{p}_n , \delta \omega_n \rbrace_{n=1}^N
\enskip . \label{state_4}
\end{equation}

The rough hard-disk system (RHDS) is an extension of the common smooth hard-disk system  (SHDS), where the 
particles rotate and exchange translational and rotational energy at a collision.
This implies a coupling between translational and rotational degrees of freedom.
The ``rough-hard-collision'' transformation rule is established from
the conservation laws for energy, linear momentum, and angular momentum.
A first version of the RHDS was already introduced in the late $19^{\textrm{th}}$ century by Bryan \cite{Bryan}.
Explicit motion equations were given by Chapman and Cowling \cite{Chapman} in 1939.
Initially, roughness was introduced as the maximum possible roughness in the model,
where a collision reverses the relative surface velocity at the point of contact of two colliding particles.
Such a model was studied by O'Dell and Berne \cite{ODell}.
Subsequently, Berne and Pangali also investigated models with partial roughness \cite{Berne, Pangali}.
Here we use the maximum-roughness model of Chapman and Cowling \cite{Chapman}.

The energy of the RHDS is purely kinetic,
\begin{displaymath}
K = \frac{1}{2}\sum_{n=1}^N \left[ \frac{\left( \vec{p}_n\right)^2}{m} + I ( \omega_n)^2 \right] \enskip ,
\end{displaymath}
where $I$ is the moment of inertia, which is taken to be identical for all disks.

We first recall the collision rules for two rough hard spheres $k$ and $\ell$ in three dimensions \cite{Chapman, vMP2009}.
Let $\vec{q} = \vec{q}_k - \vec{q}_{\ell}$ and $\delta \vec{q} = \delta \vec{q}_k - \delta \vec{q}_{\ell}$
denote the respective relative positions in the phase space and in the tangent space,
and analogously, $\vec{p} = \vec{p}_k - \vec{p}_{\ell}$ and $\delta \vec{p} = \delta \vec{p}_k - \delta \vec{p}_{\ell}$
the relative momenta.
For convenience, we define the following quantities:
$\vec{\Omega}=\vec{\omega}_k+\vec{\omega}_{\ell}$,
$\delta \vec{\Omega} = \delta\vec{\omega}_k+ \delta \vec{\omega}_{\ell}$,
the unit vector $\vec{n}=\vec{q} / \sigma$, the dimensionless moment of inertia
$\kappa = {4 I} / {m \sigma^2}$,
and the velocity of the impact point on the surface of the scatterer,
\begin{displaymath}
\vec{g} = \frac{\vec{p}}{m}+ \frac{\sigma}{2} \, ( \vec{n} \times \vec{\Omega} ) \enspace .
\end{displaymath}
During a collision, only the components associated with the colliding particles $k$ and $\ell$
of the vector ${\bf \Gamma}$ change \cite{vMP2009}.
\begin{eqnarray}
{\vec{q}}_{k}^{\;\prime} &=& \vec{q}_{k} \enspace ,
\nonumber \\
{\vec{q}}_{\ell}^{\;\prime} &=& \vec{q}_{\ell} \enspace ,
\nonumber \\
{\vec{p}_k}^{\;\prime} &=&  \vec{p}_k - m \gamma \, \vec{g} - \beta (\vec{n} \cdot \vec{p}) \, \vec{n} \enspace ,
\nonumber \\
{\vec{p}_{\ell}}^{\;\prime} &=&  \vec{p}_{\ell} + m \gamma \, \vec{g} + \beta (\vec{n} \cdot \vec{p}) \, \vec{n} \enspace ,
\nonumber \\
{\vec{\omega}_{k}}^{\;\prime} &=& \vec{\omega}_{k} + \frac{2 \beta}{\sigma} \, (\vec{n} \times \vec{g}) \enspace ,
\nonumber \\
{\vec{\omega}_{\ell}}^{\;\prime} &=& \vec{\omega}_{\ell} + \frac{2 \beta}{\sigma} \, (\vec{n} \times \vec{g}) \enspace ,
\label{colli_map_1}
\end{eqnarray}
where the prime, here and below, refers to the state immediately after the collision.
The constant parameters $\gamma={\kappa}/(1+\kappa)$ and $\beta=1/(1+\kappa)$
control the coupling between translational and rotational degrees of freedom.
If the moment of inertia vanishes ($\kappa \rightarrow 0$ and $\beta \rightarrow 1$)
the translational and rotational degrees of freedom decouple.

Using the general equation, Eq.~(14) of Ref.\cite{DPH1996}
the linearization of  Eq. (\ref{colli_map_1}) yields the collision map for the
perturbation vectors ${\bf \delta \Gamma}$ \cite{vMP2009},
\begin{eqnarray}
\delta {\vec{q}_k }^{\;\prime} &=& \delta {\vec{q}_k } + \Big[ \gamma  \vec{g} + \frac{\beta}{m}  (\vec{n} \cdot \vec{p})  
\vec{n} \Big]  \delta \tau_c \enspace , \nonumber \\
\delta {\vec{q}_{\ell} }^{\;\prime} &=& \delta {\vec{q}_{\ell} } - \Big[ \gamma  \vec{g} + \frac{\beta}{m}  (\vec{n} \cdot \vec{p})
\vec{n} \Big]  \delta \tau_c \enspace , \nonumber\\
\delta {\vec{p}_k }^{\;\prime} &=& \delta {\vec{p}_k } - m \gamma  \delta \vec{g}_c
- \frac{\beta}{\sigma}  \left[ \left( \delta \vec{q}_c \cdot \vec{p} \right)  \vec{n} +
\sigma (\vec{n} \cdot \delta \vec{p} \big)  \vec{n} + (\vec{n} \cdot \vec{p})  \delta \vec{q}_c \right] , \nonumber \\
\delta {\vec{p}_{\ell} }^{\;\prime} &=& \delta {\vec{p}_{\ell} } + m \gamma  \delta \vec{g}_c
+ \frac{\beta}{\sigma}  \left[ \left( \delta \vec{q}_c \cdot \vec{p} \right)  \vec{n} +
\sigma (\vec{n} \cdot \delta \vec{p} \big)  \vec{n} + (\vec{n} \cdot \vec{p})  \delta \vec{q}_c \right] , \nonumber \\
\delta {\vec{\omega}_{k} }^{\;\prime} &=&
\delta \vec{\omega}_{k} + \frac{2 \beta}{\sigma^2} 
\Big[ \delta \vec{q}_c \times \vec{g} + \sigma ( \vec{n} \times \delta\vec{g}_c ) \Big] \enspace , \nonumber \\
\delta {\vec{\omega}_{\ell} }^{\;\prime} &=&
\delta \vec{\omega}_{\ell} + \frac{2 \beta}{\sigma^2} 
\Big[ \delta \vec{q}_c \times \vec{g} + \sigma ( \vec{n} \times \delta\vec{g}_c ) \Big] \enspace ,
\label{colli_map_2}
\end{eqnarray}
where $\delta \tau_c = - (\delta \vec{q} \cdot \vec{n} ) / ( \vec{p}/m \cdot \vec{n} )$
is the infinitesimal time shift between the collision of the reference and the perturbed trajectories,
and $\delta \vec{q}_c =  \delta \vec{q} + \delta \tau_c \, {\vec{p}}/{m}$
corresponds to the shift (in configuration space) between the collision points of the reference and of the satellite trajectories.
Furthermore,
\begin{displaymath}
\delta \vec{g}_c  =  \frac{\delta \vec{p}}{m} +
\frac{1}{2} \Big[ \delta \vec{q}_c \times \vec{\Omega} + \sigma ( \vec{n} \times \delta \vec{\Omega} ) \Big] \enspace .
\end{displaymath}

The three-dimensional maps (\ref{colli_map_1}) and (\ref{colli_map_2}) remain valid for the case of planar
rough disks.
In this case all position and velocity vectors are placed in the $xy$-plane, and all angular velocity vectors
are perpendicular to this plane with a single non-vanishing $z$-component
(denoted $\omega_i$ for disk~$i$) \cite{vMP2009}.
Analogous definitions apply to the perturbation vectors.

For smooth hard disks an elegant matrix formulation has been introduced in Refs. \cite{WB2004,Morriss_2010} for the 
collision maps in the  phase and tangent space. It  has been generalized by us to the rough disk case. For details
we refer to Ref.  \cite{Hadrien_2011}.

The equations in this section are used in the following to construct the time evolution of the perturbation vectors
and the corresponding time-dependent local Lyapunov exponents.
The global exponents are obtained by a time average along a long trajectory.

\section{Simulation results: Covariant versus Gram-Schmidt vectors}
\label{gs_covariant}

As usual, we consider reduced units for which the particle diameter $\sigma$, the particle mass $m$ and the
kinetic energy per particle, $K/N$, are unity. Here, $K$ is the total energy, which is purely kinetic.
Lyapunov exponents are given in units of $\sqrt{K/N m \sigma^2}$.
We consider two-dimensional fluid-like systems consisting of $N=88$ hard disks at a density $\rho=0.7$.
The box with periodic boundary conditions has an aspect ratio $A=2/11$,
the same value as in our previous work on smooth hard disks~\cite{Hadrien2010_1},
to facilitate comparison.

For the computation of the covariant vectors according to the algorithm of Ginelli \textit{et al.}
\cite{Ginelli,Hadrien2010_1,Hadrien2010_2}, 
the system was first relaxed forward in time for $t_r = 10^5 \tau$ time units.
Then, the Gram-Schmidt vectors were stored for $t_{s} = 5 \cdot 10^4 \tau$ time units.
Finally, the system was iterated backward in time, and covariant vectors were computed for times in the interval
$\left[ t_r, t_r+ 0.9 \, t_{s} \right] $.
Here, $\tau=0.6$ is the time interval between two successive Gram-Schmidt re-orthonormalization steps.

For $\kappa > 0$ (three degrees of freedom per particle), the energy per particle is taken $1.5$ in our reduced units.
For $\kappa = 0$ (two degrees of freedom per particle), the energy per particle is set to unity.
With this choice, the kinetic temperatures agree and a comparison between the SHDS and RHDS is meaningful.

The numerical accuracy for the global Lyapunov exponents quoted below for this system is of the order of  0.01.
This estimate is based on the fluctuations of the GS-exponents, if the storage period $\tau_s$ is subdivided into ten
independent time intervals. The agreement between the covariant and GS-exponents is of the same order of magnitude.
Great care was taken for the choice of the simulation parameters $\tau_r$ and $\tau_s$ to allow the perturbation vectors
enough time to rotate into position. The local exponents are found to be insensitive to a doubling of these parameters. 
The symmetries discussed below for the local exponents, GS or covariant,  are typically obeyed with a numerical accuracy 
of the order of 0.001.  For systems with only three or four disks in Section \ref{null_subspace}, these accuracy 
estimates  are even better by at least a factor of two.  This rather high accuracy and fast convergence is a virtue of the 
strongly mixing hard-particle systems studied here and could not be achieved with comparable numerical effort 
for particles interacting with a smooth potential.

\begin{figure}[t]
\centering
\includegraphics[angle=0,width=1\textwidth]{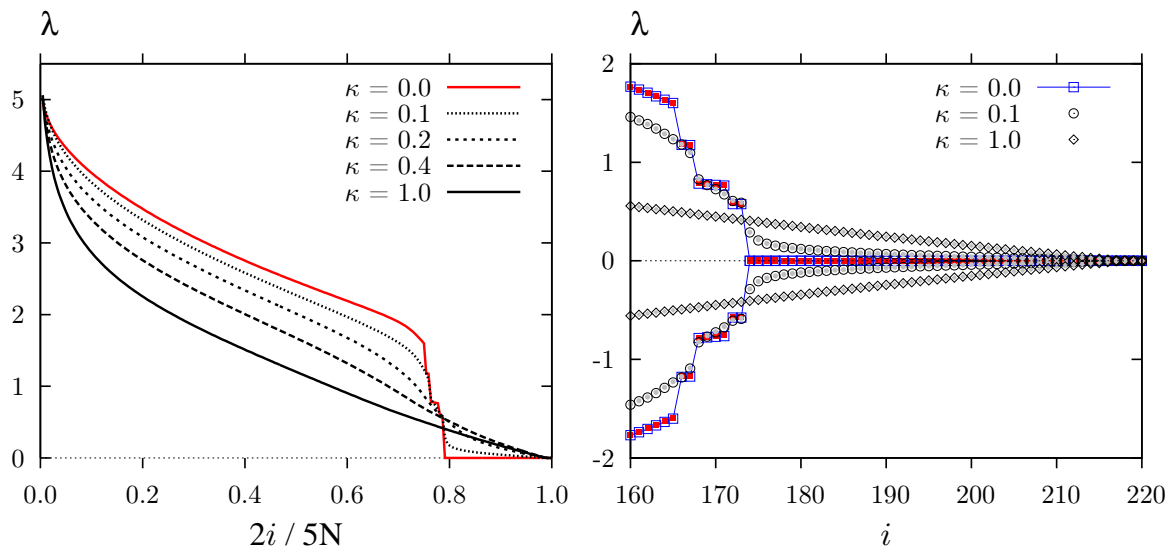}
\caption{(Color online) Left panel: Lyapunov spectra for an equilibrium system consisting of a gas of 88 rough hard disks
for various values of the coupling parameter ($\kappa \in \left\lbrace 0,0.1,0.2,0.4,1\right\rbrace $).
The spectra are calculated with the help of the covariant vectors.
Reduced indices $2i/5N$ are used on the abscissa.
Although the spectrum is defined only for integer $i$, lines are drawn for clarity.
Right panel:
Enlargement of the Lyapunov spectra for the small exponents (in absolute value).
Only the spectra for $\kappa \in \lbrace 0,\,0.1,\,1 \rbrace$ are shown.
The open symbols indicate exponents computed from the GS vectors, the full symbols exponents obtained from the
covariant vectors.
Steps  due to Lyapunov modes can be observed for $\kappa \le 0.1$, but the mode strutcure  disappears without trace for larger  $\kappa$.}
\label{spectrum}
\label{spectrum_macro}
\end{figure}

The left panel of Fig.~\ref{spectrum} shows the Lyapunov spectra computed with the covariant vectors
of the RHDS for five values of the coupling parameter $\kappa \in \left\lbrace 0,0.1,0.2,0.4,1\right\rbrace $.
The right panel of  Fig.~\ref{spectrum_macro} shows an enlargement for the smallest exponents of these spectra.
Conjugate exponent pairs are plotted with the same index $i$ on the abscissa, where now $i \in \{1, . . .,5N/2\}$.
The open symbols are computed from the GS vectors in the forward direction of time,
the full symbols from the covariant vectors during the time-reversed iteration.
Considering the size of the system ($N = 88$), the agreement is excellent.
The figure clearly displays the conjugate pairing symmetry $\lambda_i=\lambda_{5N+1-i}$ expected for time reversible energy-conserving systems.

We can see in the right panel of Fig.~\ref{spectrum_macro} the influence exerted by the moment of inertia $I$ on the spectra.
For $I=0$ corresponding to $\kappa=0$,
the steps in the spectrum due to degenerate exponents are a clear indication for the presence of Lyapunov modes.
According to the classification in \cite{Eckmann:2005}, the steps with a twofold
degeneracy are transverse ($T$) modes --~$T(1, 0)$ and $T(2,0)$ from right to left.
Similarly, the steps with a fourfold degeneracy of the exponents are longitudinal-momentum ($LP$) modes --~$LP(1, 0)$
and $LP(2,0)$ again from right to left.
Since our simulation cell is rather narrow, only modes with wave vectors $\vect{k}$
parallel to the $x$-axis of the (periodic) cell appear, leaving 0 for the second argument.
The modes for covariant vectors of smooth hard disks was studied in~\cite{Hadrien2010_1}.
As usual, ``transverse" and ``longitudinal" refer to the spatial polarization with respect to the wave vector $\vect{k}$
of the wave-like pattern.
As is evident from Fig.~\ref{spectrum_macro}, the mode structure in the spectra gradually disappears
if $\kappa$ is increased.
This is a consequence of rotation-translation coupling.

\subsection{Angles between the position and momentum parts of perturbation vectors}
\label{gs_covariant_1}

Next, we show how differently the GS and covariant vectors span the tangent space.
We consider, for each Lyapunov index $i$, the angle $\Theta_i$ between the spatial part and the translational-momentum part
of the perturbation vector $\delta {\bf \Gamma}_i$
--~either a Gram-Schmidt vector $\vect{g}_i$ or a covariant vector $\vect{v}_i$.
Defining the $2N$-dimensional position vectors
$\delta \vect{q}_i=(\delta \vec{q}_i^{\;1},\ldots,\delta \vec{q}_i^{\;N})$ and
the $2N$-dimensional momentum vectors
$\delta \vect{p}_i=(\delta \vec{p}_i^{\;1},\ldots,\delta \vec{p}_i^{\;N})$, we compute
$\cos \Theta_i =(\delta \vect{q}_i \cdot \delta \vect{p}_i)/(\Vert \delta \vect{q}_i \Vert \, \Vert \delta \vect{p}_i \Vert )$.
Note, that for the reduced units we use throughout, the positions and momenta are dimensionless.
The elements of the perturbation vectors consisting of the $N$ angular velocities of the disks are not considered.

\begin{figure}[t]
\begin{minipage}[c]{1\linewidth}
$\langle \; \cos\,  \Theta \; \rangle$ \hfill \,
\end{minipage}\\
\begin{minipage}[c]{.49\linewidth}
\includegraphics[angle=0,width=1\textwidth]{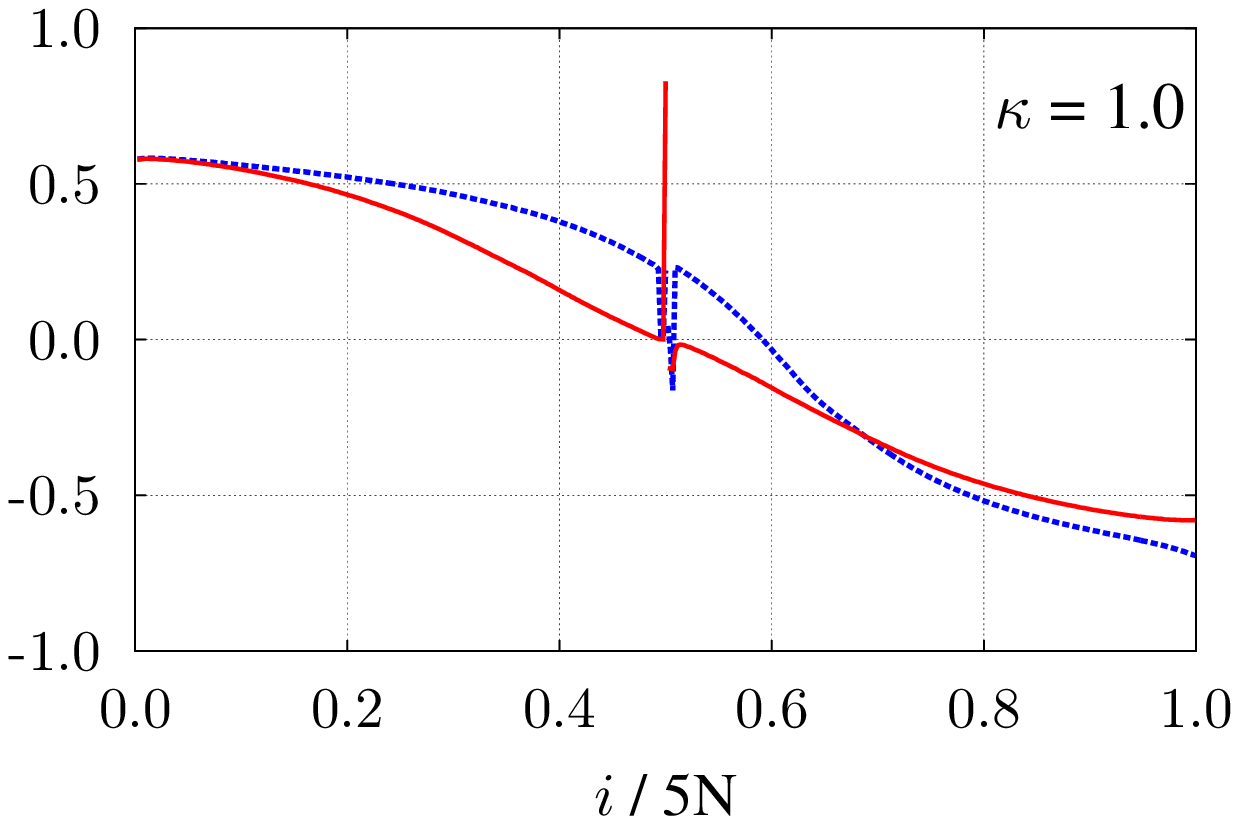}\\
\includegraphics[angle=0,width=1\textwidth]{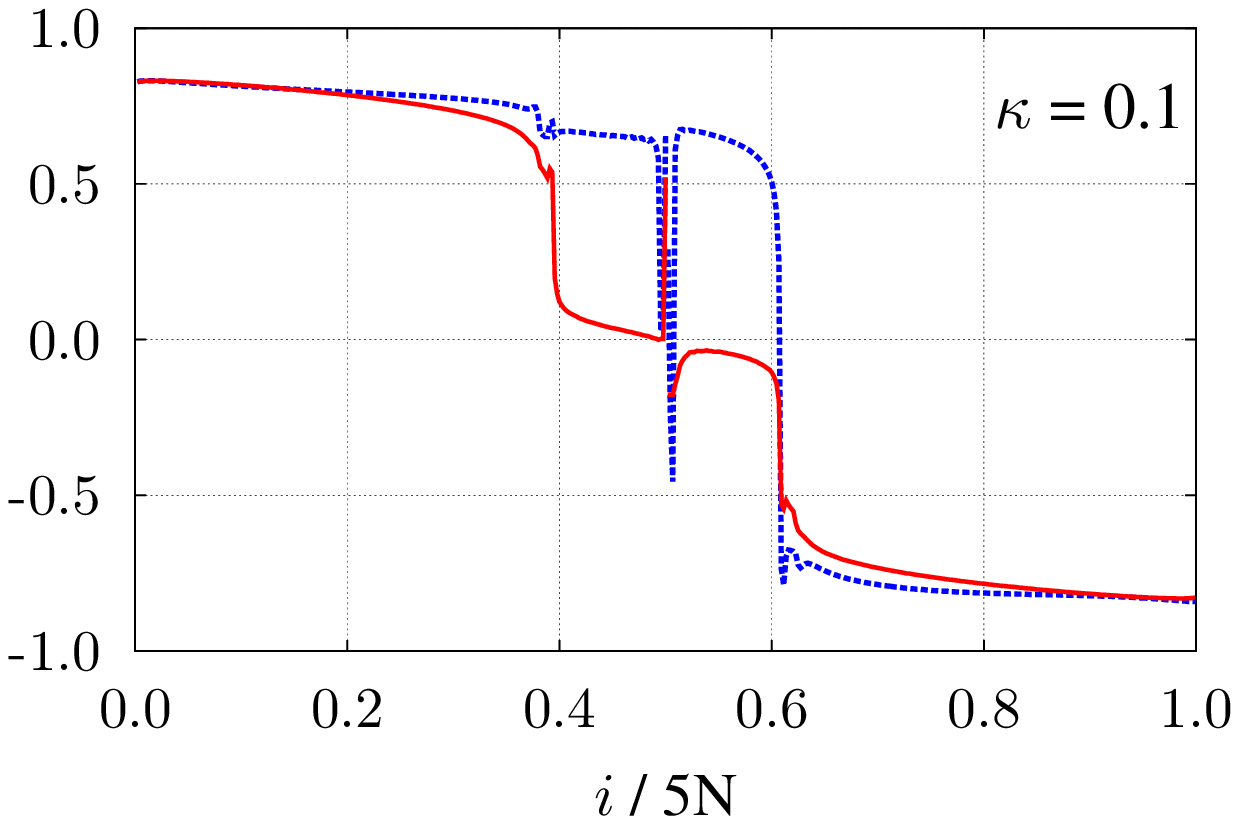}
\end{minipage} \hfil
\begin{minipage}[c]{.49\linewidth}
\includegraphics[angle=0,width=1\textwidth]{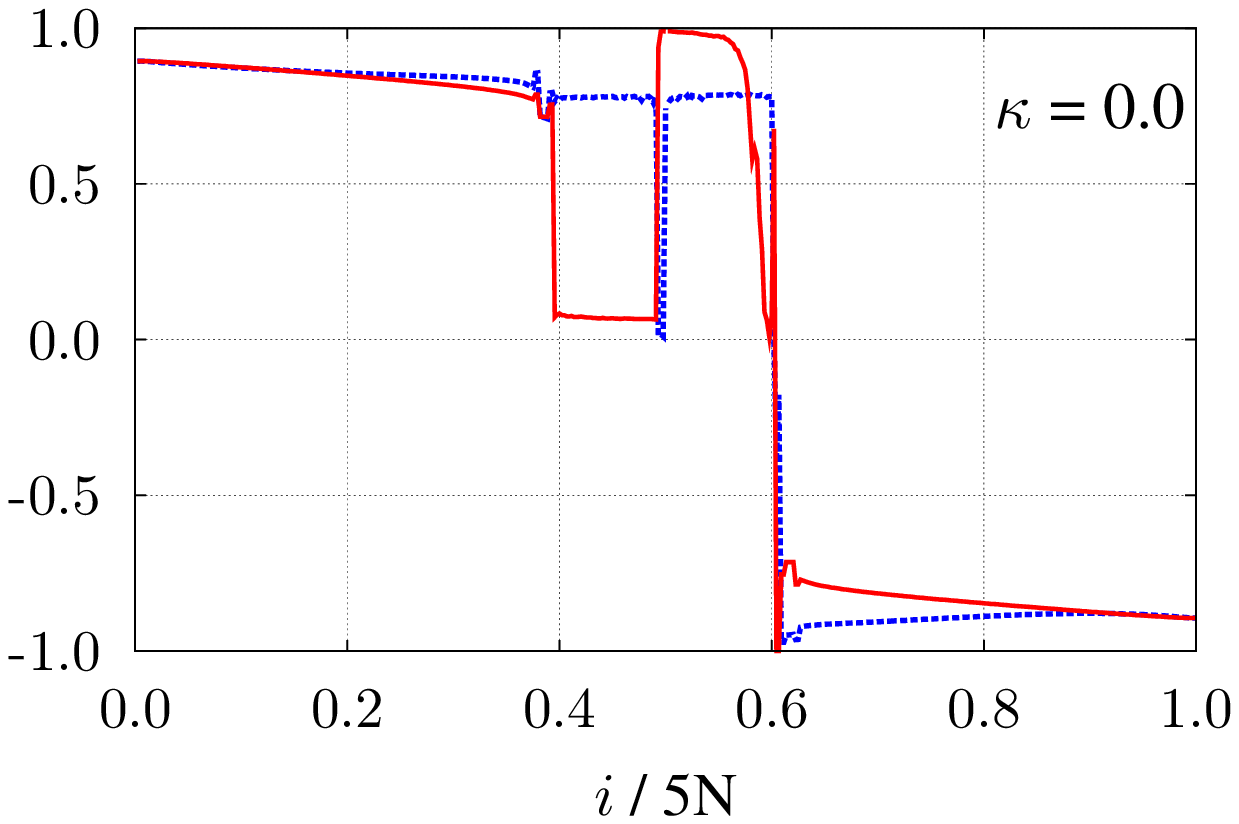}\\
\includegraphics[angle=0,width=1\textwidth]{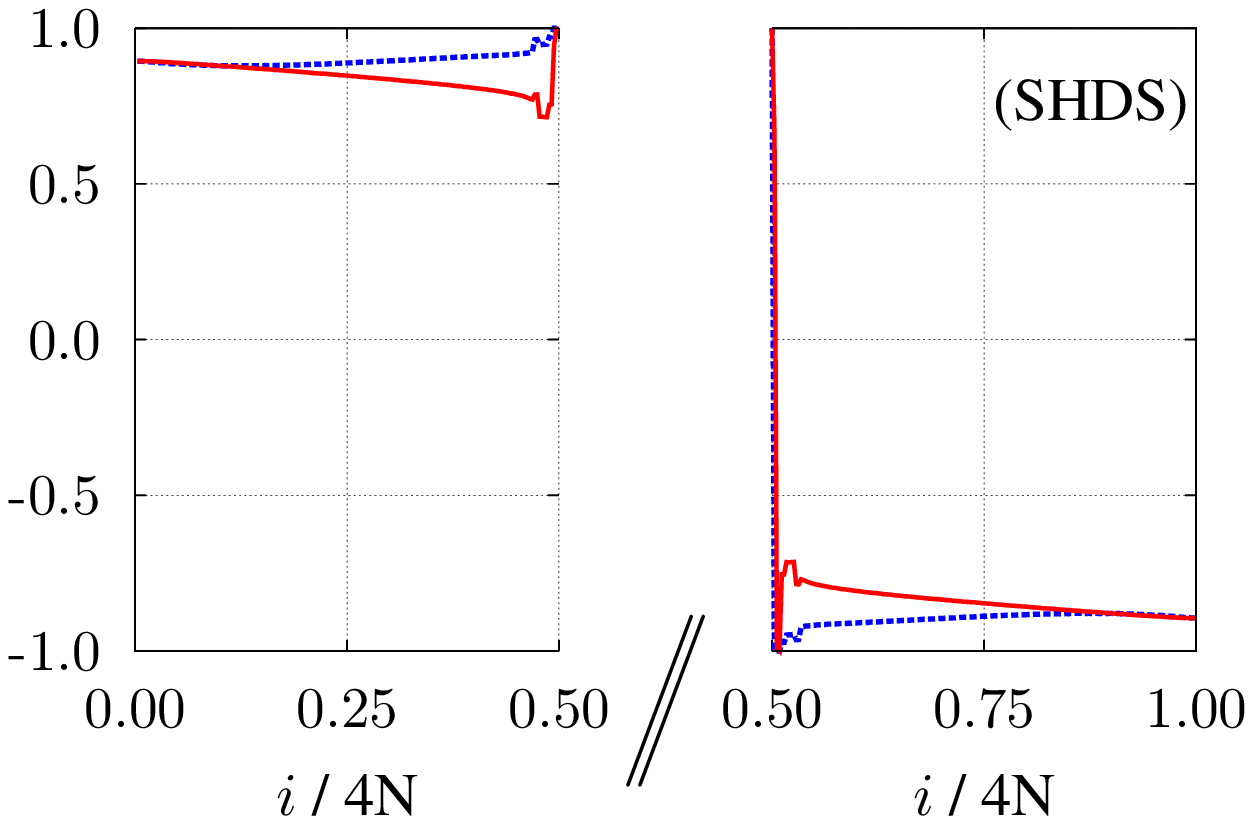}
\end{minipage}
\caption{(Color online) Time-averaged value of
$\cos(\Theta)=(\delta \vect{q} \cdot \delta \vect{p})/(\Vert \delta \vect{q}\Vert \, \Vert \delta \vect{p} \Vert )$
as a function of the Lyapunov index $i$ for a system of $N=88$ rough hard disks,
for three different values of the coupling parameter $\kappa \in \lbrace 0 , 0.1 ,1 \rbrace$,
as well as for a standard system of $N=88$ smooth hard disks.
The density is $\rho=0.7$ and the aspect ratio is $A=2/11$.
Here, $\delta \vect{q} $ and $\delta \vect{p}$ are the two $2N$-dimensional vectors of all
position perturbations and of all tranlational-momentum perturbations respectively,
for GS vectors $\vect{g}$ (blue line) and covariant vectors $\vect{v}$ (red line).
The case of smooth hard disks is treated in the bottom-right panel.}
\label{cos}
\end{figure}

In Fig.~\ref{cos}, the time average of $\cos \Theta_i$ is shown as a function of the Lyapunov index $i$
for three values of the coupling parameter, $\kappa \in \lbrace 0,\,0.1,\,1 \rbrace$.
As a reference, the respective result for the SHDS is also shown in the bottom-right panel of Fig.~\ref{cos}.
Excluding the rotation-dominated part of the spectra  $(0.4< i/5N < 0.6) $, one observes in the covariant case 
(red curves) and for $\kappa > 0$ that the positive branch is anti-symmetry with respect to
the conjugate negative branch,
\begin{displaymath}
\langle \cos \left( \Theta_{\mathtt{D}+1-i} (t) \right) \rangle = - \langle \cos \left( \Theta_{i} (t) \right) \rangle
\qquad \forall \; i \, \in \lbrace 1 , \ldots ,\mathtt{D}/2 \rbrace
\enskip .
\end{displaymath}
The covariant vectors evolve according to the natural tangent bundle
and obey the time-reversal symmetry prescribed by the respective dynamical rules.
That means that the $i^{\,\textrm{th}}$ expanding vector $\vect{v}_i$
in the time-forward direction becomes the conjugate contracting vector $-\vect{v}_{\mathtt{D}+1-i}$ in the time-backward direction.
The $\left\langle \cos \Theta \right\rangle$ for the GS vectors (blue curves) does not show this time-reversal symmetry.
Whereas the deviations for $\kappa = 0.1$ outside the range $0.4 \leq \, i/5N \, \leq 0.6$ are comparatively small
(bottom-left panel of Fig.~\ref{cos}),
they are pronounced for $\kappa = 1$ (top-left panel of Fig.~\ref{cos}).
This is to be expected due to the fact that the GS vectors do not obey time-reversal symmetry in general as outlined
in detail in~\cite{Hadrien2010_2}.


It is interesting to note that the simple SHDS (bottom-right panel in  Fig.~\ref{cos})
shows anti-symmetrical behavior of $\left\langle \cos \Theta \right\rangle$  both for the covariant vectors (as expected)
and for the GS vectors.  
For the latter, this can only be a consequence of the symplectic symmetry of the GS vectors for Hamiltonian systems
\cite{Hadrien2010_2, Ramaswamy}.
This was explicitly stated already in \cite{TM2003a},
\begin{displaymath}
\cos \left( \Theta_{\mathtt{D}+1-i} (t) \right) = - \cos \left( \Theta_{i} (t) \right)
\qquad \forall \; i \, \in \lbrace 1 , \ldots ,\mathtt{D}/2 \rbrace
\enskip ,
\end{displaymath}
for any instant of time $t$
(which is a direct consequence of Eq.~(\ref{symplectic_GS}) in Sec.~\ref{discussion}).
Of course, also the time-averaged spectra are anti-symmetric as displayed in the bottom-right panel of
Fig.~\ref{cos}.
However, the rough-hard-disk model for $\kappa = 0$, keeping the angular velocity variables included in the
GS re-orthonormalization process, does not show this symmetry (top-right panel of Fig.~\ref{cos}).
This is not a numerical artifact since the GS vectors, which do not display the expected symmetry,
are used in the construction of the covariant vectors, which display the symmetry.
Thus, it is not sufficient to simply put $\kappa$ to 0 for rough disks to generate the same tangent space dynamics as found for SHDS.
It is also necessary to explicitly exclude the redundant angular velocity components of the perturbation vectors from the
GS re-orthonormalization procedure.

It is well known that the Lyapunov vectors associated with the large positive and negative Lyapunov exponents are strongly
localized in physical space \cite{HBP98,TM2003b,FHPH2004}. We have verified this property also for the rough disks.  
For details we refer to Ref. \cite{Hadrien_2011}. Not too surprisingly, the localization is even stronger for the covariant vectors,
whose directions in tangent space are solely determined by the tangent flow and are not affected by re-orthonormalization constraints.
The localization of a covariant vector with a large positive Lyapunov exponent  is found to agree with the localization  of its conjugate 
twin with negative exponent  \cite{Hadrien_2011}. For orthogonal Gram-Schmidt vectors this symmetry does not exist. 
 
\section{Null subspace and vanishing exponents for the RHDS}\label{null_subspace}

\begin{figure}[t]
\centering
\begin{minipage}[r]{0.9\linewidth}
\begin{tabular}{c | c  c}
$N=3$
& \multicolumn{1}{c|}{\begin{footnotesize}including the orientation\end{footnotesize}}
& {\begin{footnotesize}excluding the orientation\end{footnotesize}}\\
\hline
{\rotatebox{90}{\begin{footnotesize} periodic boundaries\end{footnotesize}}}
&\includegraphics[angle=0,width=0.45\textwidth]{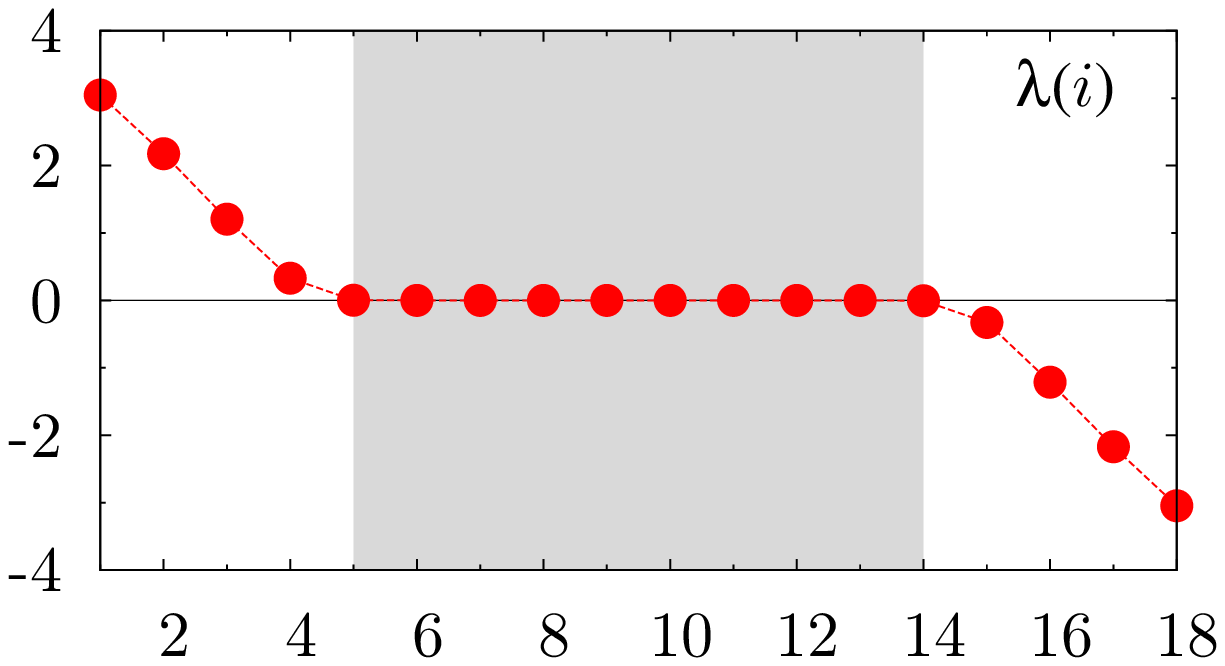}
&\includegraphics[angle=0,width=0.45\textwidth]{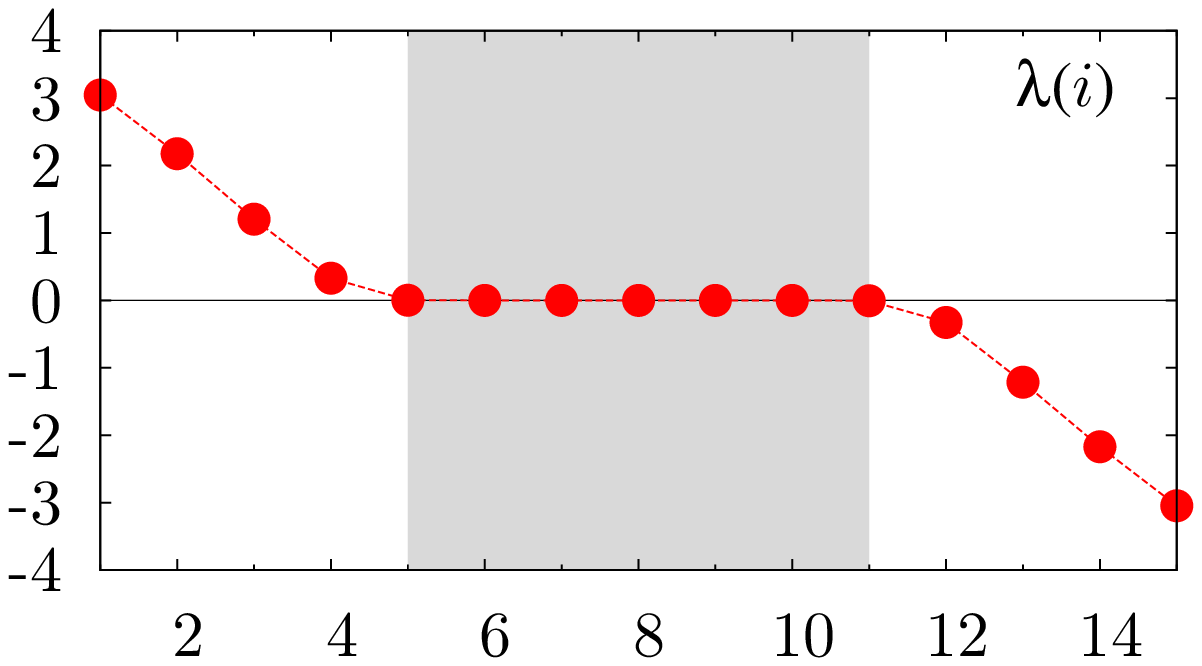}\\
\cline{1-1}
{\rotatebox{90}{\begin{footnotesize} reflecting boundaries\end{footnotesize}}}
&\includegraphics[angle=0,width=0.45\textwidth]{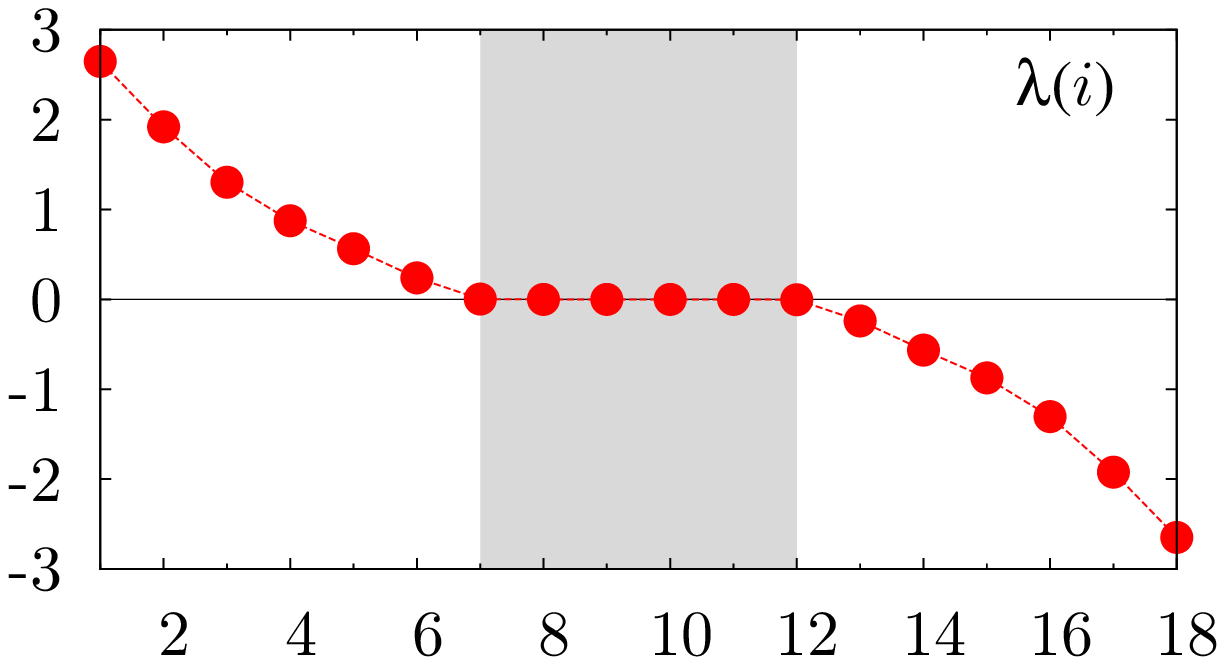}
&\includegraphics[angle=0,width=0.45\textwidth]{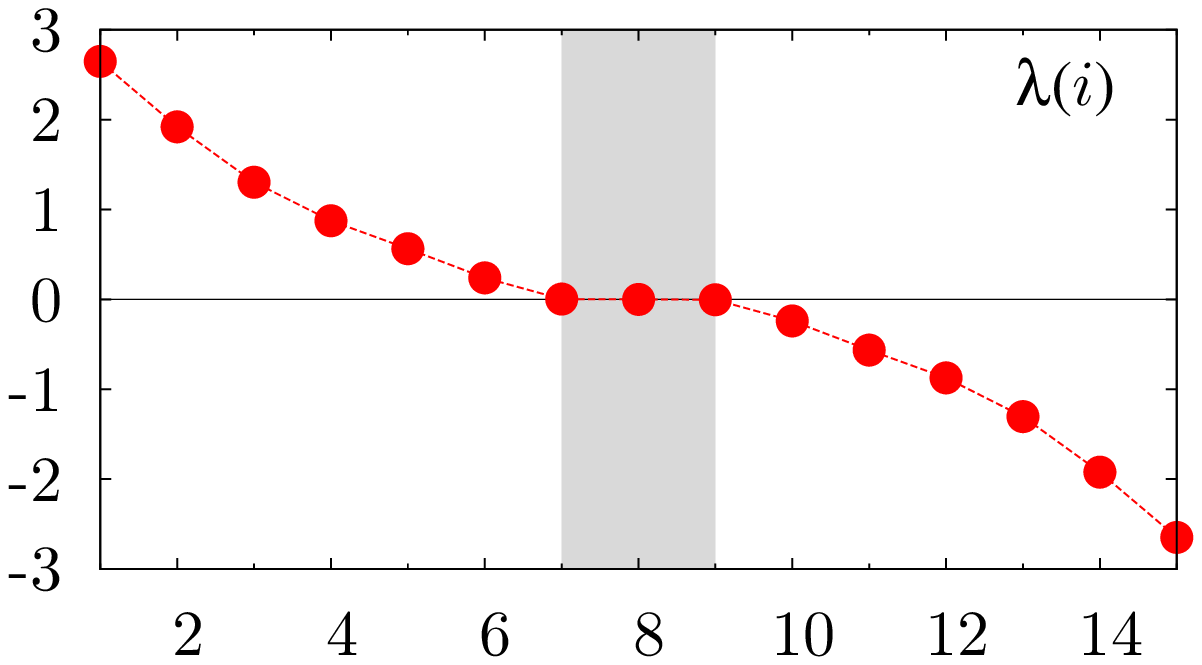}
\end{tabular}
\caption{(Color online) Lyapunov spectrum of the RHDS, with $N=3$ disks,
for the cases of periodic and reflecting boundary conditions,
where the orientation angles of the disks are either included or excluded from the phase-space and tangent-space dynamics.
$\kappa = 0.5$.}
\label{null_space_odd}
\end{minipage}\\
\vspace{5mm}
\begin{minipage}[r]{0.9\linewidth}
\begin{tabular}{c | c  c}
$N=4$
& \multicolumn{1}{c|}{\begin{footnotesize}including the orientation\end{footnotesize}}
& {\begin{footnotesize}excluding the orientation\end{footnotesize}}\\
\hline
{\rotatebox{90}{\begin{footnotesize} periodic boundaries\end{footnotesize}}}
&\includegraphics[angle=0,width=0.45\textwidth]{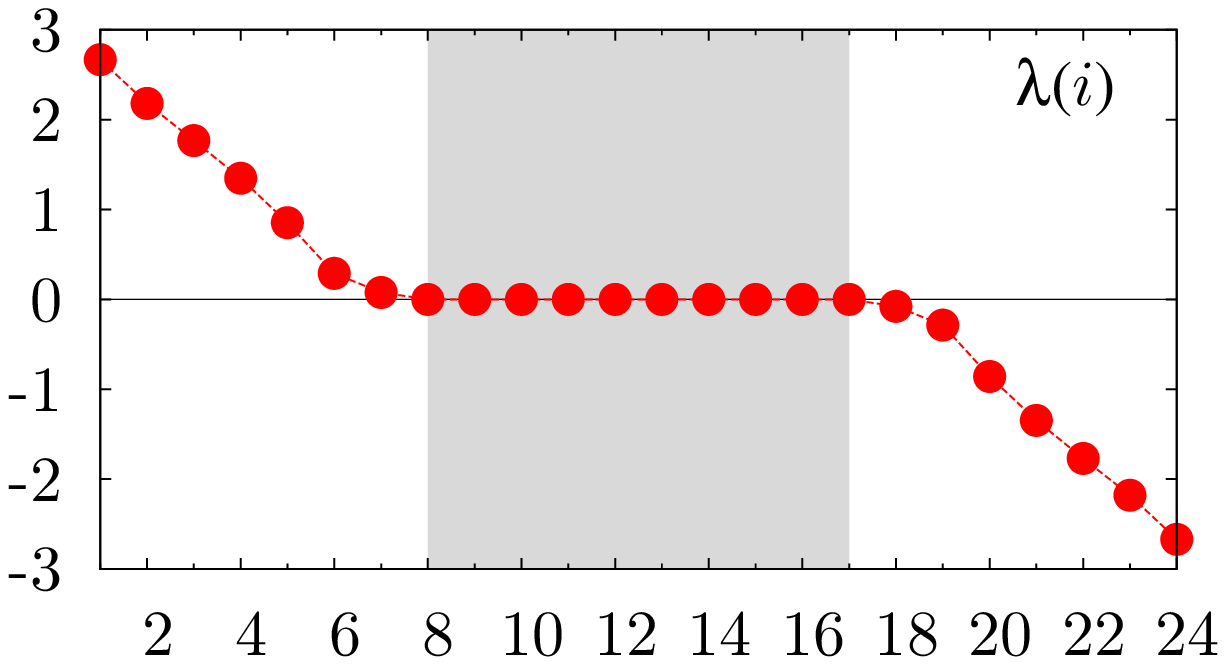}
&\includegraphics[angle=0,width=0.45\textwidth]{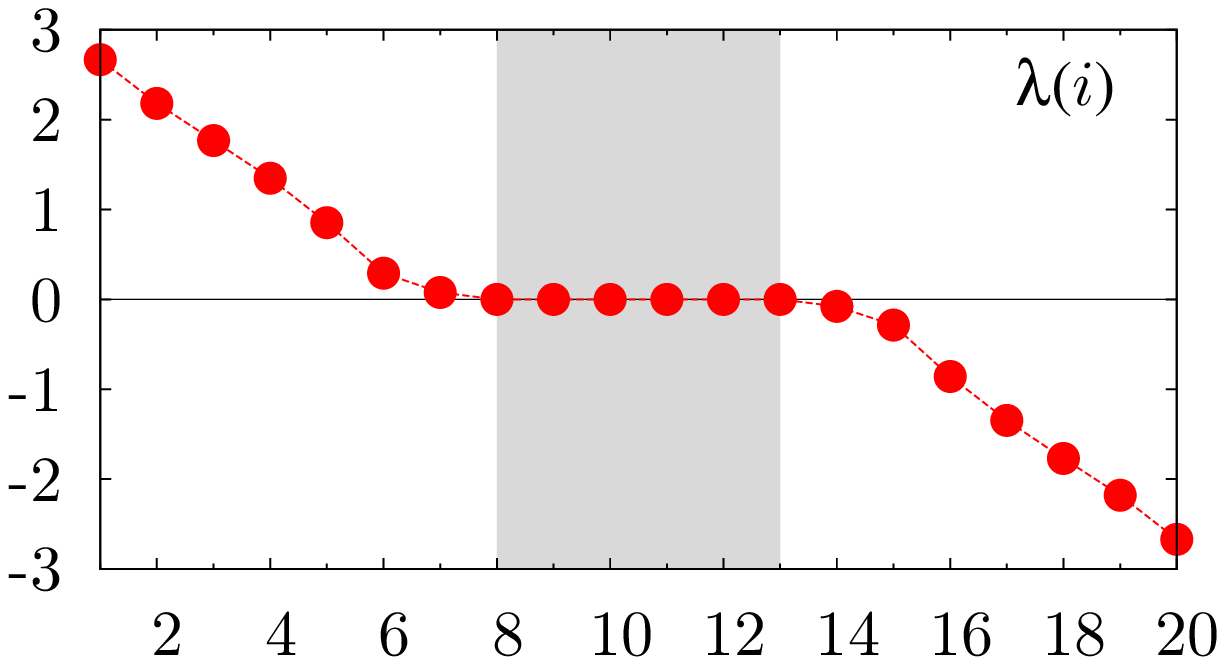}\\
\cline{1-1}
{\rotatebox{90}{\begin{footnotesize} reflecting boundaries\end{footnotesize}}}
&\includegraphics[angle=0,width=0.45\textwidth]{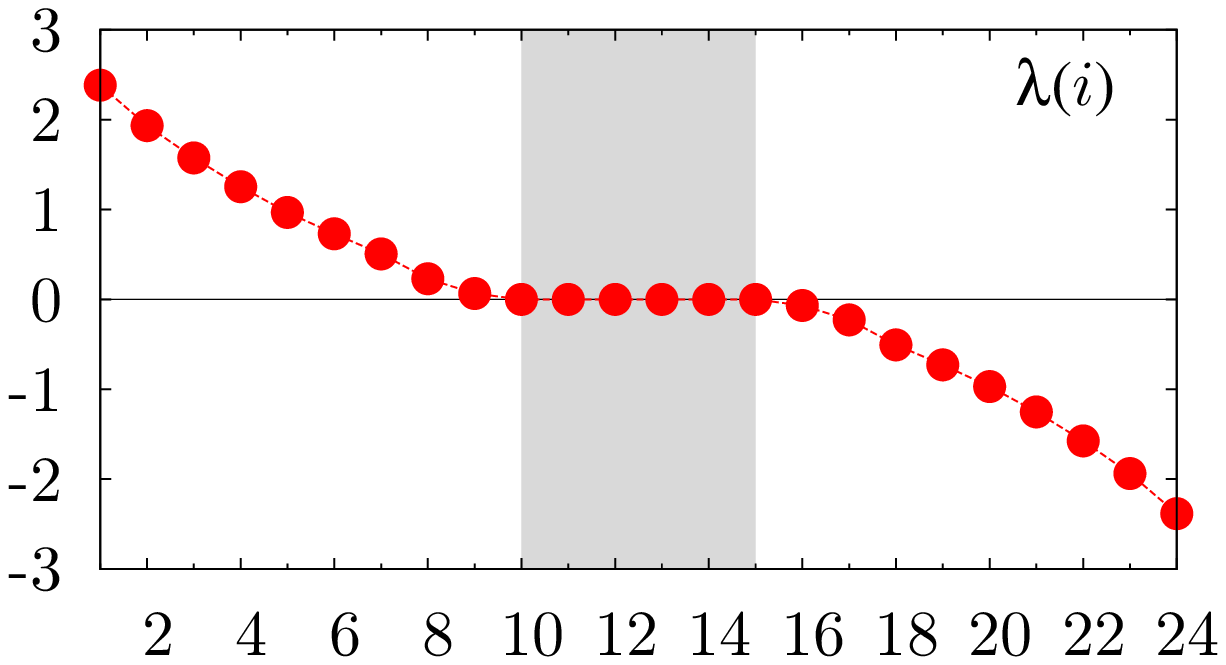}
&\includegraphics[angle=0,width=0.45\textwidth]{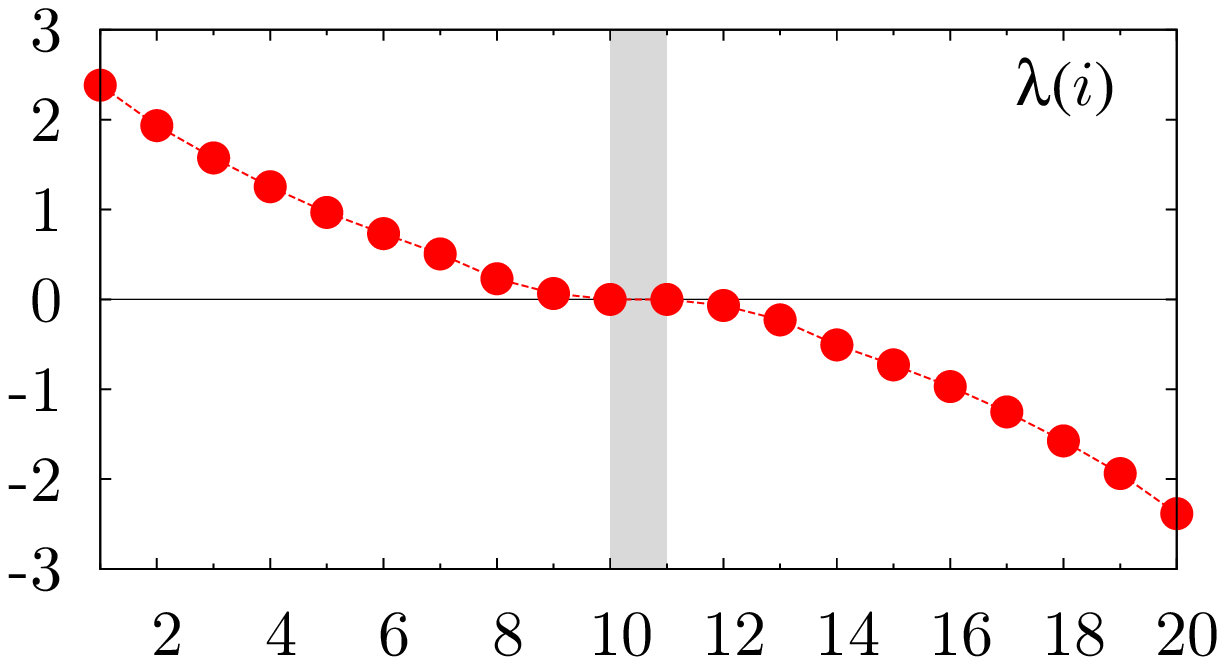}
\end{tabular}
\caption{(Color online) The same as Fig.~\ref{null_space_odd}, with $N=4$ disks.}
\label{null_space_even}
\end{minipage}
\end{figure}

The subspace of the tangent space associated with the vanishing exponents is of special interest.
Let us consider some experimental results first.
In Fig.~\ref{null_space_odd}
the Lyapunov spectra of a rough-hard-disk system consisting of an odd number of particles, namely $N=3$ disks, are shown.
The panels in the top row are for periodic boundary conditions, which we exclusively used so far.
The panels in the bottom row, however, are for reflecting boundary conditions, for which the tangent vectors change according
to Eq.~(20) in~\cite{DPH1996} for a reflection at the boundary.
The essential difference is that translational momentum conservation does not apply in this case (both in $x$- and $y$-directions),
which results in a reduction of the vanishing exponents by 4 with respect to the periodic-boundary case.
The panels on the left-hand side of Fig.~\ref{null_space_odd}
include the orientation angles $\lbrace \theta_n \rbrace_{n=1}^N$ in the list of independent variables in Eq.~(\ref{state_2})
and the respective perturbations $\lbrace \delta \theta_n \rbrace_{n=1}^N$ in Eq.~(\ref{state_4}).
Although the  $\lbrace \theta_n \rbrace_{n=1}^N$ do not affect the dynamics in the phase space at all
(as discussed further below),
the associated perturbations $\lbrace \delta \theta_n \rbrace_{n=1}^N$
contribute to the evolution of the tangent bundle since they take part in the GS re-orthonormalization.
At the collision of particles $k$ and $\ell$ these perturbation components change ~\cite{DPH1996} according to
\begin{equation}
\left. 
\begin{array}{l}
\delta \theta_{\ell}^{\;\prime} =  \delta \theta_{\ell} - \frac{2 \beta}{\sigma} \; \left( \vec{n} \times \vec{g} \right) \delta \tau_c \\
\delta \theta_{k}^{\;\prime} = \delta \theta_{k} - \frac{2 \beta}{\sigma} \; \left( \vec{n} \times \vec{g} \right) \delta \tau_c \;,
\end{array}
\quad
\right\rbrace
\label{orientation}
\end{equation}
which augment the Eqs.~(\ref{colli_map_2}) in this case.
The panels on the right-hand side of Fig.~\ref{null_space_odd}
are for the standard representation of the RHSD without the explicit consideration of the disk orientations
$\lbrace \theta_n \rbrace_{n=1}^N$ (see Eqs.~(\ref{colli_map_1}) and~(\ref{colli_map_2})).
The spectra in the analogous Fig.~\ref{null_space_even} are for an even number of disks, $N=4$.

The case of an even number of disks in Fig.~\ref{null_space_even} is well understood 
whether the orientation angles are explicitly considered or not. In all cases, two of the exponents vanish 
due to the invariance with respect to time translation and energy conservation. For reflecting boundaries and excluding 
orientations (bottom-right panel) these are the only symmetries leading only to two vanishing exponents. For periodic 
boundaries and orientations excluded
(top-right panel), four other vanishing exponents need to be added due to  the invariance with respect to a uniform 
translation of the origin in the  $x$- and $y$ directions and to a uniform shift of the translational momenta  in the
$x$- and $y$ directions. Also for the reflecting boundary case with orientations explicitly included (bottom-left panel), six
exponents vanish, but for a different reason. The four additional vanishing exponents are due to the invariance 
with respect to a rotation of any of the four disks by an arbitrary angle. Finally, in the periodic boundary case with
explicit consideration of the angles (top-left panel) all symmetries mentioned above contribute to increase the  total number of 
vanishing exponents to  $6+N$, where $N = 4$ in Fig.~\ref{null_space_even}.

For an odd number of disks the situation is not clear to us. We still do not have a satisfactory explanation for the
experimental result that an odd number of vanishing exponents is found, if the orientations of the disks are 
excluded from the considerations (see the panels on the right-hand side of Fig.~\ref{null_space_odd}). 
Similarly,  if the orientation angles are included, a naive consideration of intrinsic symmetries would suggest 5 and 9
vanishing exponents for reflecting and periodic boundaries, respectively. Instead, the experimental results shown in the 
left-hand side panels of Fig.~\ref{null_space_odd} yield 6 and 10 vanishing exponents,  one more than expected.
One additional symmetry seems to exist, which is not properly accounted for in this case. 

\section[Transversality versus order violation in the local Lyapunov spectrum]
{Transversality versus order violation in the local Lyapunov spectrum}\label{transversality}

The dynamics of the tangent bundle is characterized by a complete set of dissociated sub-bundles,
which are defined by the Oseledec splitting,
\begin{displaymath}
\mathbf{TX}_{{\bf \Gamma}}=
\vect{E^u} \oplus \; {\cal N} \oplus \vect{E^s}
\enskip ,
\end{displaymath}
at any point ${\bf \Gamma}$ in phase space.
Here, $\vect{E^u}= \vect{v}_{1} \oplus \cdots \oplus \vect{v}_{\mathtt{D}/2-3}$ and
$\vect{E^s}=\vect{v}_{\mathtt{D}/2+4} \oplus \cdots \oplus \vect{v}_{\mathtt{D}}$ are the covariant unstable and stable subspaces, 
respectively, and ${\cal N}$ is the covariant null subspace
(the dimension $\mathtt{D}$ of the phase space equals $5N$ for RHDS and $4N$ for SHDS).
The multiplicative ergodic theorem provides us with the following inequalities for the global Lyapunov spectrum
\begin{equation}
\lambda_1 \ge \cdots \ge \lambda_{\mathtt{D}/2-3} > \left[ \lambda^{(0)} \right] > \lambda_{\mathtt{D}/2+4}
\ge \cdots \ge \lambda_{\mathtt{D}}
\label{lyapunov_order}
\enskip ,
\end{equation}
where the equal sign relates to degenerate exponents, and
$\left[\lambda^0\right] = 0$ is sixfold degenerate
(assuming energy conservation and momentum conservation in $x$- and $y$-directions).
These inequalities, which guarantee a hierarchy of the instabilities,
imply also a certain dynamical isolation in the tangent bundle.

The order in Eq.~(\ref{lyapunov_order}) may be different,
if the local Lyapunov exponents are averaged over a finite time (instead of an infinite time for the global exponents)
to which we refer as ``loss of isolation''.
It may be expressed in terms of the order violation probability for the finite-time averaged local exponents, $\overline{\Lambda}^{\,\tau}_j$,
which is the probability that $\overline{\Lambda}^{\,\tau}_j < \overline{\Lambda}^{\,\tau}_i$, whereas for
the global exponents $\lambda_j \geqslant \lambda_i$ holds. Therefore, a spectrum is not violated, if all probabilities
$P(\overline{\Lambda}^{\,\tau}_j < \overline{\Lambda}^{\,\tau}_i)=0$
for all times $\tau>\tau_c$, where $\tau_c>0$.
It implies that the angles between the Oseledec subspaces are bounded away from zero \cite{YR2008},
which indicates uniform transversality for the respective subspaces.
These considerations may be applied to the angles between all possible covariant vector pairs.

\begin{figure}[p]
\centering
\begin{minipage}[c]{.65\linewidth}
\includegraphics[width=1\textwidth]{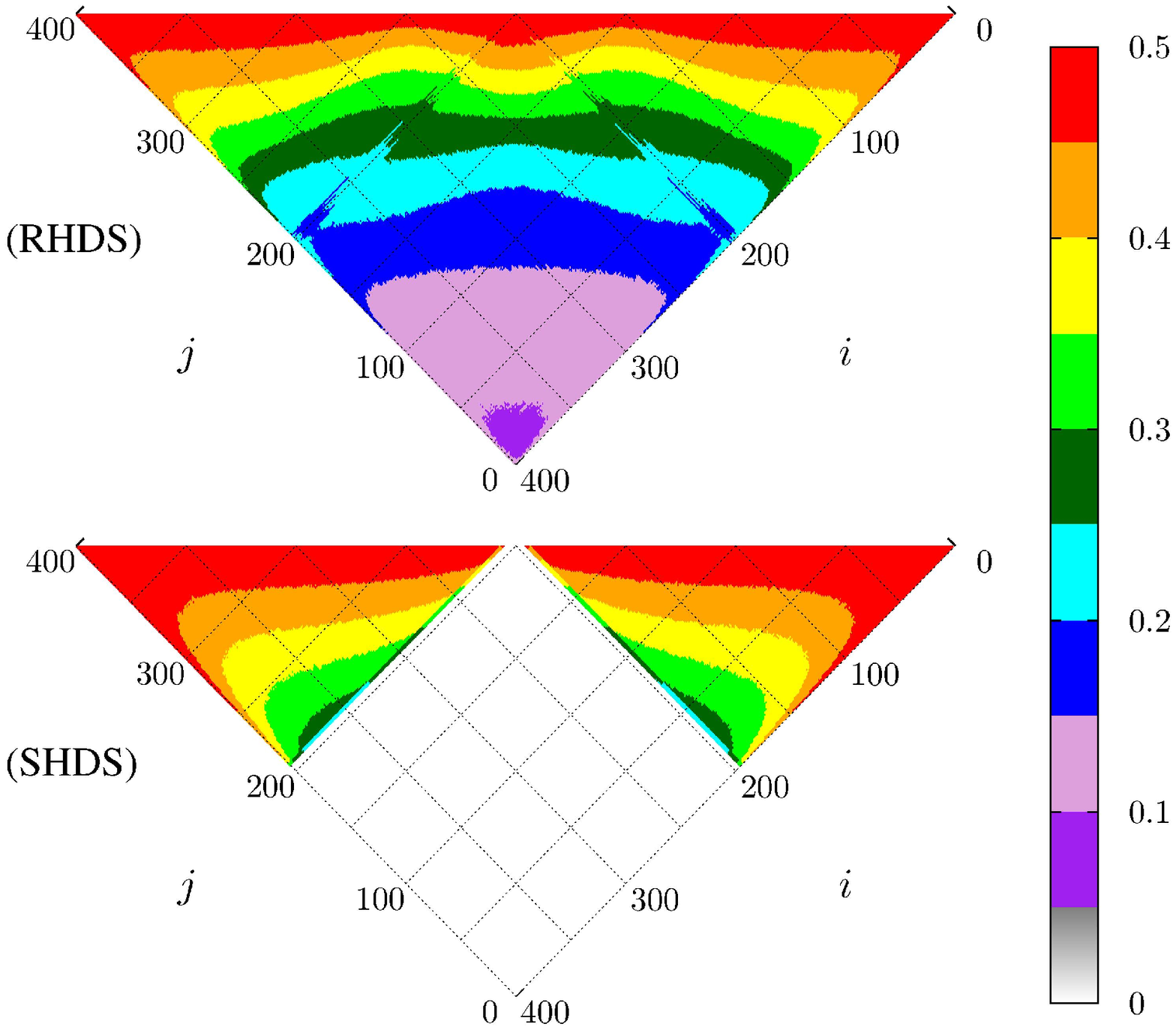}
\end{minipage}
\begin{minipage}[c]{.05\linewidth}
\flushleft{\rotatebox{90}{
\footnotesize{$P \left( ^{\textrm{cov}}{\overline{\Lambda}}^{\;0.01}_{\,j} < \, ^{\textrm{cov}}{\overline{\Lambda}}^{\;0.01}_{\,i} \right) $}
}}\end{minipage}
\caption{(Color online) Violation probability contour plot for covariant local exponents for averaging times $\tau=0.01$.
Top panel: rough-hard-disk system ($\kappa = 0.1$). Bottom panel: smooth-hard-disk system.
The order violation probability
$P ( ^{\textrm{cov}}{\overline{\Lambda}}^{\;0.01}_{\,j} < \, ^{\textrm{cov}}{\overline{\Lambda}}^{\;0.01}_{\,i} ) $
is shown for all Lyapunov index pairs $(i,j)$ for which the global exponents obey $\lambda_j \geqslant \lambda_i$ (such that $j<i$).}
\label{violation}
\vspace{8mm}
\begin{minipage}[c]{.65\linewidth}
\includegraphics[width=1\textwidth]{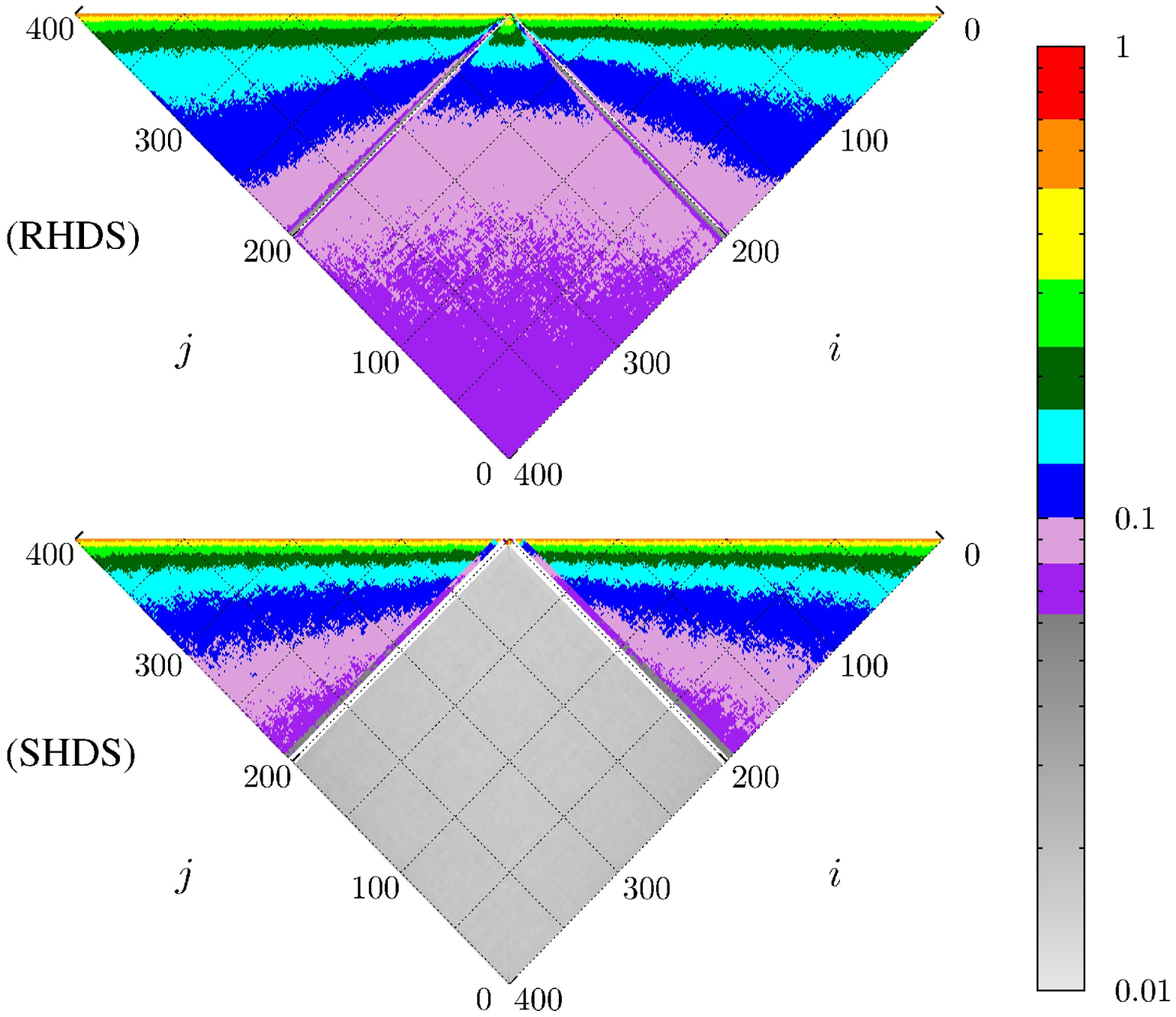}
\end{minipage}
\begin{minipage}[c]{.05\linewidth}
\flushleft{\rotatebox{90}{
\small{$ \left\langle \, \vert \vect{v}_j \cdot \vect{v}_i \vert \, \right\rangle $}
}}\end{minipage}
\caption{(Color online) Contour plot of $\langle \vert \vect{v}_j \cdot \vect{v}_i \vert \rangle$
for all Lyapunov index pairs $(i,j)$ such that $j<i$.
Top panel: rough-hard-disk system ($\kappa = 0.1$). Bottom panel: smooth-hard-disk system.}
\label{scalar_plot}
\end{figure}

In Fig.~\ref{violation} we show contour plots of 
$P ( ^{\textrm{cov}}{\overline{\Lambda}}^{\;\tau}_{\,j} < \, ^{\textrm{cov}}{\overline{\Lambda}}^{\;\tau}_{\,i} ) $, $\tau=0.01$,
for all $j<i$ (provided $\lambda_j \geqslant \lambda_i$), and compare results for rough hard disks (RHDS) with parameters
$N=80$, $\kappa=0.1$ and $\rho=0.7$, with analogous results for smooth hard disks with parameters 
$N=100$ and $\rho=0.7$.
The averaging period $\tau=0.01$ is small enough such that time averaging does not affect the local exponents much,
$^{\textrm{cov}}{\overline{\Lambda}}^{\;0.01}_{\,i} \approx \,^{\textrm{cov}}{\Lambda}_{\,i} $, for all $i$.
$10^5$ such intervals were used for the construction of each contour plot.
The regime $UU$, for which $i\in [2,187]$ and  $j\in [1,i-1]$, refers to vector pairs from the unstable manifold and is located
in a triangle in the upper right corner of these plots.
The regime $SS$, for which $i\in [205,400]$ and  $j\in [204,i-1]$, refers to vector pairs from the stable manifold and is located
in a triangle in the upper left corner of these plots.
The regime $SU$, for which $i\in [204,400]$ and  $j\in [1,187]$, refers to the case where $\vect{v}_i$
is a covariant vector from the stable manifold and $\vect{v}_j$ from the unstable manifold.
This regime is located in a square in the center of the plot, which separates $SS$ from $UU$.
Vector pairs involving at least one vector from the central manifold are located in narrow stripes in between these domains.
A fundamental difference between the rough and smooth disk case may be observed.

For smooth disks (SHDS) in Fig.~\ref{violation}, the violation probability $P$ vanishes for all vector pairs from $SU$,
which confirms our previous results that the minimum angle between the stable and unstable subspaces is well bounded away from
zero, at least for finite $N$~\cite{Hadrien2010_1}.
An entanglement of covariant vectors and, hence, a loss of isolation of covariant subspaces due to an order violation of local
exponents only occurs within the stable and unstable manifolds.
This causes the finite numbers for the violation probability $P$ in $UU$ and $SS$.
Our result also implies that the stable and unstable manifolds have the same dimension at all points in phase space.

For rough disks (RHDS), the violation probability varies rather smoothly in Fig.~\ref{violation}
and does not vanish even for vector pairs from the central manifold. There is evidence that
stable covariant vectors may be  ``entangled'' with unstable covariant vectors. It has been argued that such an entanglement may be 
responsible for the disappearance of Lyapunov modes in particle systems \cite{YR2008}. We suspect that one reason for this
disappearance  may be  the non-orthogonality of the central manifold to the stable and unstable manifold
for the RHDS.

Related and complementary information may be gained by considering the scalar products between covariant vector pairs.
In Fig.~\ref{scalar_plot} we show
-- for the same systems as before --
contour plots of $\langle \vert \vect{v}_j \cdot \vect{v}_i \vert \rangle$ for all $j<i$.
The same domains $UU$, $SS$ and $SU$ may be distinguished.

Let us first again consider the SHDS first (bottom panel in Fig.~\ref{scalar_plot}).
Most remarkably, vector pairs from $SU$, which correspond to the uniformly shaded square, are not orthogonal to each other.
This confirms our result of Ref. \cite{Hadrien2010_1}, where we found that the stable and unstable manifolds are transverse but
not orthogonal. The central manifold, however, is orthogonal to both the stable and unstable manifolds. 

\begin{figure}[t]
\centering
\includegraphics[angle=0,width=0.6\textwidth]{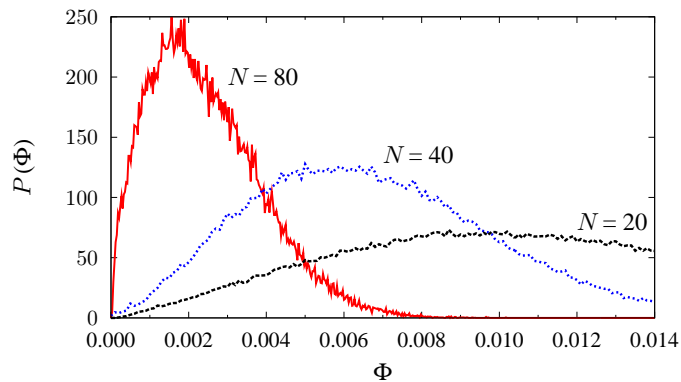}
\caption{Probability distribution for the minimum angle $\Phi$ between the stable and unstable manifolds
for three rough hard disk systems  of size $N$ as indicated by the labels. The coupling parameter $\kappa$ equals 0.4.} 
\label{hyperbolicity}
\end{figure}

The situation is slightly more complicated for the RHDS  (top panel in Fig.~\ref{scalar_plot}).
One observes a very smooth variation of  $\langle \vert \vect{v}_j \cdot \vect{v}_i \vert \rangle$
for all three regimes and deviations from orthogonality everywhere.
Most notably, these deviations persist also between a vector from the central manifold and another vector from
the stable or unstable manifolds (the narrow stripes between $SU$ and $UU$ respective $SS$).
To check for transversality, we show in  Fig. \ref{hyperbolicity}
 the probability distribution for the minimum angle $\Phi$ between the stable subspace $\vect{E^u}$ and the unstable subspace
 $ \vect{E^s}$. It is computed from the smallest principal angle
between the two subspaces \cite{Kuptsov,Bjoerck,Knyazev}. If the covariant vectors belonging to 
 ${\vect E}^u$  and   $\vect E^s $ are arranged as the column vectors of matrices 
 ${\vect V^u}$ and ${\vect V^s}$, respectively, the QR decompositions ${\vect V^u} = {\bf Q^u} {\bf R^u} $ 
 and  ${\vect V^s} = {\bf Q^s} {\bf R^s} $ provide matrices ${\bf Q^u}$ and ${\bf Q^s}$, with which 
 the matrix ${\bf M} = \{{\bf Q^u}\}^T {\bf Q^s}$ is constructed. The singular values of ${\bf M}$ are
 equal to the cosines of the principal angles, of which $\Phi$ is the minimum angle. 
 Fig. \ref{hyperbolicity} shows that this angle is small but positive even for the largest system considered here.
 The distributions seem to be bounded away from zero and indicate transversality for the respective subspaces. 
We conclude that for finite $N$ the rough hard-disk systems are also hyperbolic, although just about. 

\section{The RHDS is not symplectic}\label{discussion}

It is easily seen from the motion equations that the SHDS and  RHDS conserve energy. As a consequence,
the phase volume is conserved for both systems, and
\begin{equation}
\sum_{i=1}^{\mathtt{D}} \Lambda^{GS}_i  ({\bf \Gamma}) = 0
\enskip
\end{equation}
at any point in phase space. However, this is not sufficient to ensure that they are both symplectic.
Let $\mathcal{J}$ denote the even-dimensional skew-symmetric matrix,
\begin{displaymath}
\mathcal{J} =
\left( 
\begin{array}{cc}
\mathbf{0} & \mathbf{I} \\
-\mathbf{I} & \mathbf{0} 
\end{array}
\right) 
\enskip .
\end{displaymath}
For Hamiltonian systems it was shown by Meyer \cite{Meyer1986} that for any Gram-Schmidt vector
$\vect{g}_i$ also the vector $\mathcal{J} \vect{g}_i$
belongs to the set of Gram-Schmidt vectors such that
\numparts
\begin{eqnarray}
\Lambda^{GS}_{\mathtt{D} +1-i} ({\bf \Gamma}) &=& - \Lambda^{GS}_i ({\bf \Gamma})
\label{symplectic_exponents}
\enskip ,\\
\vect{g}_{\mathtt{D} +1-i} ({\bf \Gamma}) &=& \pm \mathcal{J} \vect{g}_i ({\bf \Gamma})
\label{symplectic_GS}
\enskip ,
\end{eqnarray}
\endnumparts
for $i \in \lbrace 1, \ldots , \mathtt{D}/2\rbrace$.
In \cite{Hadrien2010_2} we have referred to Eq.~(\ref{symplectic_exponents})
as symplectic local pairing symmetry.
\begin{figure}[t]
\begin{minipage}[c]{.49\linewidth}
\includegraphics[angle=0,width=1\textwidth]{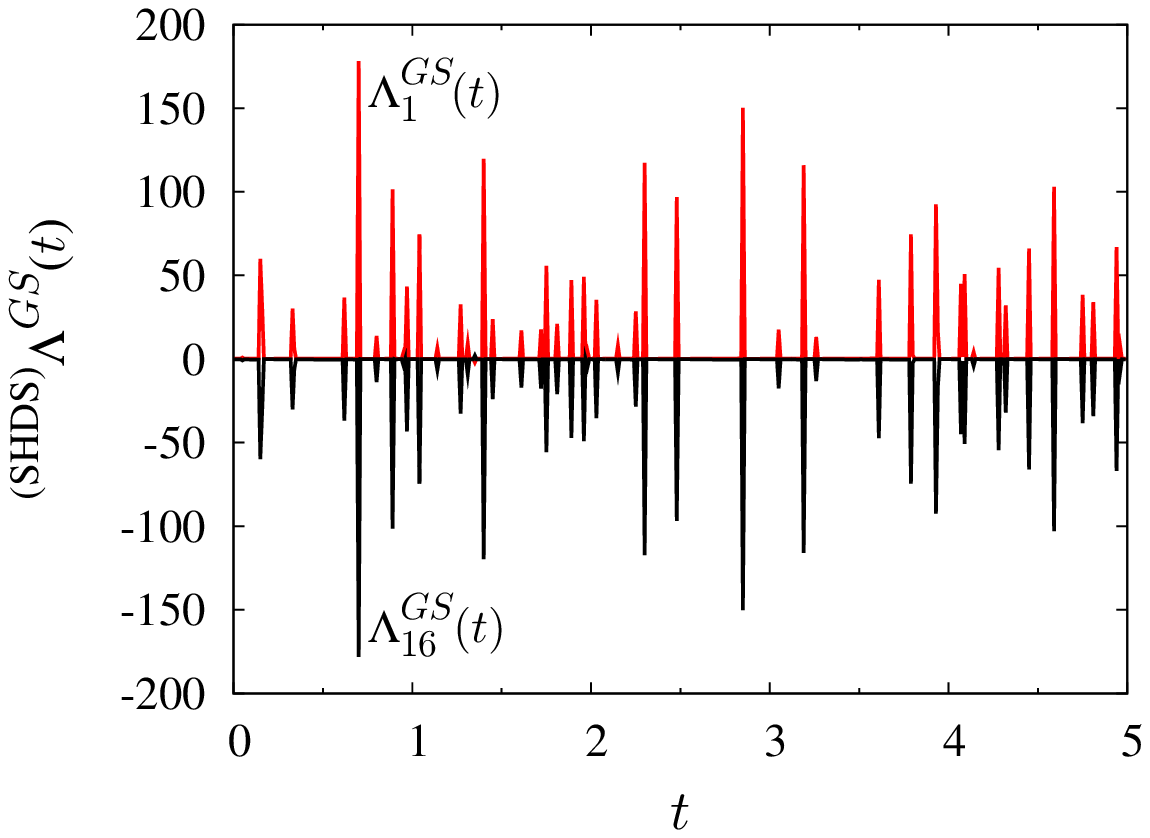}
\end{minipage} \hfill
\begin{minipage}[c]{.49\linewidth}
\includegraphics[angle=0,width=1\textwidth]{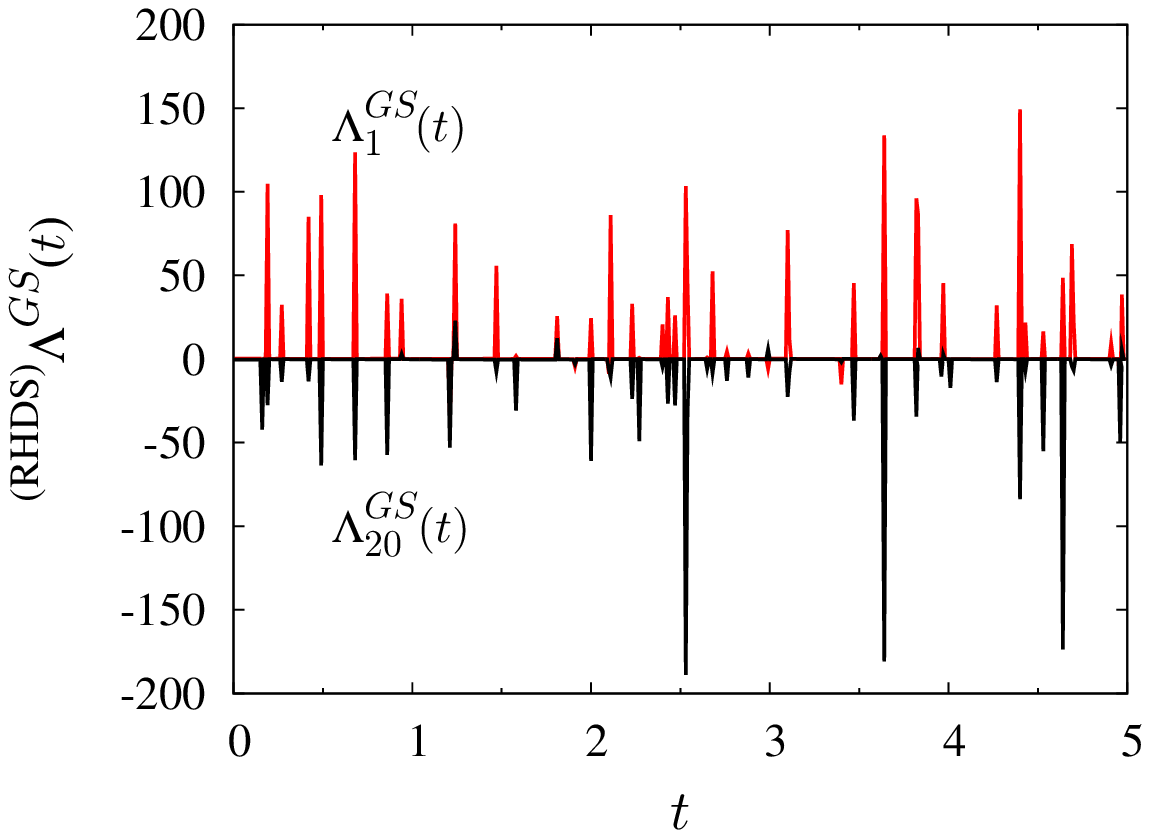}
\end{minipage}
\caption{(Color online) Test of the symplectic symmetry of Eq.~(\ref{symplectic_exponents}) for a SHDS with $N=4$ (left panel)
and for a RHDS with $N=4$ (right panel).}
\label{symplectic_exponents_test}
\end{figure}
Here, we test both expressions for the SHDS and RHDS.
In Fig.~\ref{symplectic_exponents_test} the maximum and minimum local GS exponents are shown for a SHDS (left panel)
and a RHDS (right panel). One observes that Eq.~(\ref{symplectic_exponents}) is obeyed for the smooth hard disks,
\begin{displaymath}
^{\left( \small{\textrm{SHDS}} \right)}\Lambda^{GS}_{\mathtt{D} +1-i} (t)
\, +\, ^{\left( \small{\textrm{SHDS}} \right)}\Lambda^{GS}_i (t) = 0
\enskip ,
\end{displaymath}
but not for the rough hard disks,
\begin{displaymath}
^{\left( \small{\textrm{RHDS}} \right)}\Lambda^{GS}_{\mathtt{D} +1-i} (t)
\, +\, ^{\left( \small{\textrm{RHDS}} \right)}\Lambda^{GS}_i (t) \neq 0
\enskip .
\end{displaymath}
A direct inspection of the perturbation vectors also reveals analogous results for 
Eq. (\ref{symplectic_GS}) (not shown).  
This numerical comparison demonstrates that the RHDS is not a Hamiltonian system
as already suspected in Sec.~\ref{gs_covariant}.

The non-Hamiltonian character of the RHDS  may be shown also analytically
by explicitly considering the transformation rules in the tangent bundle
\cite{Arnold_1,Arnold_2,Frankel}. 
Any linear transformation $\mathcal{S}$ is symplectic
if, and only if, it satisfies the identity
\begin{equation}
\mathcal{S}\,^{\dagger}
\mathcal{J} \;
\mathcal{S}=
\mathcal{J}
\enskip .
\label{identity}
\end{equation}
 Rewriting the linearized collision map (such as Eq.(\ref{colli_map_2}) for the SPHS)
in matrix form,
\begin{equation} 
 \delta {\bf  \Gamma}^{\;\prime} = \mathcal{S}\;  \delta {\bf  \Gamma},
 \label{sym}
\end{equation}
we examined the validity of  (\ref{identity}) for the respective
smooth- and rough-hard-disk models \cite{Hadrien_2011}. 
After some tedious calculation  (omitted here)  it is  shown 
that Eq. (\ref{identity}) holds for the SHDS, but fails 
for the rough disks, whether the disk orientations are 
explicitly considered or eliminated.  

\section{Discussion and summary} \label{summary}

We  study smooth (SHDS) and rough hard-disk systems (RHDS)
in two dimensions, which resemble  dense planar gases in thermodynamic equilibrium.
The investigation of the tangent-space structure for both models reveal strong qualitative differences. 
Our approach is based on the computation of the Oseledec splitting of the tangent space into a  hierarchy of
covariant subspaces spanned by the individual covariant Lyapunov vectors, which are associated with
respective global (time-averaged) and local (time-dependent) Lyapunov exponents. Wherever
meaningful, we also compare the properties of the orthonormal Gram-Schmidt (GS) vectors with those
of the  covariant vectors. Both vector sets span the tangent space at any space point. Whereas the
former constitute an orthonormal  set, the latter generally do not. However, the covariant vectors 
reflect the time-reversal symmetry of the motion equations \cite{Hadrien2010_2} and for that reason
are physically more relevant than the orthonormal GS vectors \cite{HR2010,P2011}, which are a convenient 
mathematical tool for the computation of the global Lyapunov exponents. 

There is a subtle difficulty with the equations of motion for the RHDS, whether or not the disk
orientations, which do not affect the phase-space trajectory at all, are explicitly excluded or included
for the computation of the tangent-space dynamics (Sec. \ref{null_subspace}). Exclusion and inclusion
requires phase-space (and tangent-space) dimensions of $5N$ and $6N$, respectively, where $N$ is 
the number of disks.  Whereas our main results concerning the tangent space are unaffected by this 
choice, the number of vanishing exponents (equal to the dimension of the central manifold) is strongly affected.
This number is determined by the continuous symmetries of the system and -- if the
disk orientations  are explicitly considered -- by the invariance of the motion equations with respect to
an arbitrary rotation of {\em any disk}, resulting in  $N$ additional vanishing exponents. For an 
even number of particles, the number of vanishing exponents is well understood, regardless whether
periodic or reflecting boundary conditions are considered (Fig. \ref{null_space_even}). For an odd number of particles, however,
for which our experimental results are summarized in Fig. \ref{null_space_odd}, these considerations
do not give the observed numbers. This is a vexing problem we still do not understand.     

Lyapunov modes, which are most familiar for smooth hard particle systems \cite{Posch:2000,Eckmann:2005},
exist for the RHDS only in the limit of week-coupling between translational and rotational degrees of freedom
(Sec. \ref{gs_covariant}).  Already for rather small coupling constants respective moments if inertia, the 
step structure in the Lyapunov spectrum, which is  due to the degeneracy of the Lyapunov exponents associated 
with the modes, disappears.

Another observation concerns the localization of the perturbation vectors in physical space. Whereas the
localization measure for conjugate covariant exponents coincides, it differs for the GS vectors (not shown, see \cite{Hadrien_2011} ).
This is a consequence of the time-reversal symmetry displayed by the covariant vectors.

As for the smooth hard disks  \cite{Hadrien2010_1}, the unstable, stable and central manifolds of the RHDS
are transverse to each other. But the minimal angle between the stable and unstable manifolds typically is
very small for the RHDS. For the SHDS, the central manifold is strictly orthogonal to the 
unstable and stable subspaces, which themselves are almost -- but not strictly -- orthogonal to each other. 
For the RHDS, the scalar product between two arbitrary covariant vectors does not vanish on average, but progressively 
increases, the closer the indices of the two vectors are. But it does not reach unity and, hence, no  tangencies occur even for
vector pairs with neighboring indices, at least for finite $N$ (see Fig. \ref{scalar_plot}).
If the global exponents in the  Lyapunov spectrum are ordered according to size, the  probability for this order  to be violated
by a pair of local exponents is always positive for the RHDS.  The violation probability also increases progressively,
the closer the indices of the exponent pair become, and may reach values around 0.5 for neighboring exponents
(see Fig. \ref{violation}). For the SHDS, however, this probability
vanishes, if one of the vectors is chosen from the stable, and the other from the unstable manifold.

 Finally, we numerically demonstrate that conjugate Gram-Schmidt vectors for the RHDS do not exhibit
 symplectic symmetry, whereas conjugate GS vectors for the SHDS do. This shows that the symplectic structure
 is destroyed for the rough hard disks due to the coupling of translational and rotational degrees of freedom. This interesting
 result is also confirmed by direct analytical computation.

\section{Acknowledgement}
We acknowledge support by the Austrian Science Fund (FWF),  grants P15348 and P18798-N20.



\end{document}